\documentclass[10pt,letterpaper]{article}
\usepackage[top=0.85in,left=2.75in,footskip=0.75in]{geometry}

\usepackage{changepage}

\usepackage[utf8]{inputenc}

\usepackage{textcomp,marvosym}

\usepackage{fixltx2e}

\usepackage{amsmath,amssymb}

\usepackage{cite}

\usepackage{nameref,hyperref}

\usepackage[right]{lineno}

\usepackage{microtype}
\DisableLigatures[f]{encoding = *, family = * }

\usepackage{rotating}

\raggedright
\usepackage[aboveskip=1pt,labelfont=bf,labelsep=period,justification=raggedright,singlelinecheck=off]{caption}

\makeatletter
\renewcommand{\@biblabel}[1]{\quad#1.}
\makeatother

\date{}

\usepackage{lastpage,fancyhdr,graphicx}
\usepackage{epstopdf}
\fancyheadoffset[L]{2.25in}
\fancyfootoffset[L]{2.25in}
\usepackage{amsthm, bm, xfrac}
\usepackage{graphicx, color}
\usepackage{algorithm2e}
\usepackage{multirow}

\newcommand{\indicator}{\ensuremath{\mathbf{1}}}
\newcommand{\dif}{\ensuremath{\;\mathrm{d}}}
\newcommand{\host}{\ensuremath{\text{host}}}
\newcommand{\parent}{\ensuremath{\text{parent}}}
\newcommand{\children}{\ensuremath{\text{children}}}
\newcommand{\first}{\ensuremath{\text{first}}}

\newtheorem{theorem}{Theorem}
\newtheorem{lemma}{Lemma}

\graphicspath{{figures/}}

\begin{document}
\vspace*{0.35in}

\begin{flushleft}
{\Large
\textbf\newline{Molecular Infectious Disease Epidemiology: Survival Analysis and Algorithms Linking Phylogenies to Transmission Trees}
}
\newline
\\
Eben Kenah\textsuperscript{1, 5, \textcurrency, *},
Tom Britton\textsuperscript{2},
M. Elizabeth Halloran\textsuperscript{3, 4, 5},
Ira M. Longini, Jr.\textsuperscript{1, 5}
\\
\bigskip
\bf{1} Biostatistics Department and Emerging Pathogens Institute, University of Florida, Gainesville, Florida, USA
\\
\bf{2} Department of Mathematics, Stockholm University, Stockholm, Sweden
\\
\bf{3} Vaccine and Infectious Diseases Division, Fred Hutchinson Cancer Research Center, Seattle, Washington, USA
\\
\bf{4} Department of Biostatistics, University of Washington, Seattle, Washington, USA

\bf{5} Center for Inference and Dynamics of Infectious Diseases, Fred Hutchinson Cancer Research Center, Seattle, Washington, USA
\bigskip

\textcurrency University of Florida, 22 Buckman Drive, Dauer Hall, PO Box 117450, Gainesville, FL 32611-7450

* ekenah@ufl.edu

\end{flushleft}

\section*{Abstract}
Recent work has attempted to use whole-genome sequence data from pathogens to reconstruct the transmission trees linking infectors and infectees in outbreaks. However, transmission trees from one outbreak do not generalize to future outbreaks. Reconstruction of transmission trees is most useful to public health if it leads to generalizable scientific insights about disease transmission. In a survival analysis framework, estimation of transmission parameters is based on sums or averages over the possible transmission trees. A phylogeny can increase the precision of these estimates by providing partial information about who infected whom. The leaves of the phylogeny represent sampled pathogens, which have known hosts. The interior nodes represent common ancestors of sampled pathogens, which have unknown hosts. Starting from assumptions about disease biology and epidemiologic study design, we prove that there is a one-to-one correspondence between the possible assignments of interior node hosts and the transmission trees simultaneously consistent with the phylogeny and the epidemiologic data on person, place, and time. We develop algorithms to enumerate these transmission trees and show these can be used to calculate likelihoods that incorporate both epidemiologic data and a phylogeny. A simulation study confirms that this leads to more efficient estimates of hazard ratios for infectiousness and baseline hazards of infectious contact, and we use these methods to analyze data from a foot-and-mouth disease virus outbreak in the United Kingdom in 2001. These results demonstrate the importance of data on individuals who escape infection, which is often overlooked. The combination of survival analysis and algorithms linking phylogenies to transmission trees is a rigorous but flexible statistical foundation for molecular infectious disease epidemiology.

\section*{Author Summary}
Recent work has attempted to use whole-genome sequence data from pathogens to reconstruct the transmission trees linking infectors and infectees in outbreaks. However, transmission trees from one outbreak do not generalize to future outbreaks. Reconstruction of transmission trees is most useful to public health if it leads to generalizable scientific insights about disease transmission. Accurate estimates of transmission parameters can help identify risk factors for transmission and aid the design and evaluation of public health interventions for emerging infections. Using statistical methods for time-to-event data (survival analysis), estimation of transmission parameters is based on sums or averages over the possible transmission trees. By providing partial information about who infected whom, a pathogen phylogeny can reduce the set of possible transmission trees and increase the precision of transmission parameter estimates. We derive algorithms that enumerate the transmission trees consistent with a pathogen phylogeny and epidemiologic data, show how to calculate likelihoods for transmission data with a phylogeny, and apply these methods to a foot and mouth disease outbreak in the United Kingdom in 2001. These methods will allow pathogen genetic sequences to be incorporated into the analysis of outbreak investigations, vaccine trials, and other studies of infectious disease transmission.


\section*{Introduction}
Genetic sequences from pathogen samples are an increasingly important source of information in infectious disease epidemiology. The structure of a pathogen phylogeny can reflect immunological strain selection, epidemic dynamics, and patterns of spatial spread~\cite{Grenfell2004, Wilson2005, Lemey2009}. Phylogenies linking pathogen genetic sequences sampled at known times and places have been used to investigate the origins and spread of HIV-1~\cite{Rambaut2001, Gilbert2007} and the global circulation of seasonal influenza~\cite{Nelson2007, Rambaut2008, Russell2008, Bedford2015}. Using coalescent models, phylogenies can be used to reconstruct the history of effective viral population sizes~\cite{PybusRambaut2001} or estimate basic reproductive numbers ($R_0$)~\cite{Pybus2001}. These methods can reveal details of the large-scale spread of infection that would be difficult or impossible to detect otherwise. For example, Biek~\emph{et al.}~\cite{Biek2007} showed that the invasion of the eastern United States by a raccoon-specific rabies virus occurred in seven distinct clades (each representing a different direction of spread from the epizootic origin in West Virginia) and three waves of expansion (1987--1993, 1986--1990, and 1990--1992).

When epidemic models are integrated with population genetics, several complications arise in the interpretation of the effective number of infections and the scaling of time~\cite{Volz2009, FrostVolz2010, KoelleRasmussen2011}. Recently, methods have been developed that use phylogenetic trees to make inferences about the prevalence of infection over time while accounting for epidemic dynamics~\cite{Volz2009, Stadler2013}, and some of these can incorporate time series data on the number of cases~\cite{Rasmussen2011, Rasmussen2014}. These phylodynamic methods generally assume a sparse sample of pathogen genetic sequences from a large infected population. In this paper, we consider the use of densely-sampled pathogen genetic sequences to make inferences about person-to-person transmission. 

\subsection*{Reconstructing transmission trees with genetic sequences}
The earliest use of phylogenetics in infectious disease epidemiology was to confirm or rule out a suspected source of the human immunodeficiency virus (HIV). Phylogenetic analyses were used to confirm that five HIV patients were infected at a dental practice in Florida between 1987 and 1989~\cite{Ou1992} and to rule out infection of a Baltimore patient in 1985 by an HIV-positive surgeon~\cite{Holmes1993}. A more ambitious use of phylogenetics is to reconstruct a transmission tree, which is a directed graph with an edge from node $i$ to node $j$ if person $i$ infected person $j$. An analysis by Leitner~\emph{et al.}~\cite{Leitner1996, Leitner1999} of an HIV-1 transmission cluster in Sweden from the early 1980s compared reconstructed phylogenies based on HIV genetic sequences to a true phylogeny based on the known transmission tree, times of transmission, and times of sequence sampling. The reconstructed phylogenies accurately reflected the topology of the true phylogeny, and the accuracy increased when sequences from different regions of the HIV genome were combined.

The increasing availability of whole-genome sequence data has renewed interest in combining pathogen genetic sequence data with epidemiologic data to reconstruct transmission trees. One approach to this problem is to reconstruct the transmission tree using genetic distances. Spada~\emph{et al.}~\cite{Spada2004} reconstructed the transmission tree linking five children infected with hepatitis C virus (HCV) by finding the spanning tree linking the HCV genetic sequences that minimized the sum of the genetic distances across its edges, excluding edges inconsistent with the epidemiologic data. The \emph{SeqTrack} algorithm of Jombart~\emph{et al.}~\cite{Jombart2011} generalizes this approach. It constructs a transmission tree by finding the spanning tree linking the sampled sequences that minimizes (or maximizes) a set of edge weights. Snitkin~\emph{et al.}~\cite{Snitkin2012} used this algorithm to investigate a 2011 outbreak of carbapenem-resistant \emph{Klebsiella pneumoniae} in the NIH Clinical Center, penalizing edges with large genetic distances, between patients who did not overlap in the same ward, or that required a long silent colonization. Wertheim~\emph{et al.}~\cite{Wertheim2011} constructed a network among HIV patients in San Diego by linking individuals whose sequences were $< 1\%$ distant. This was used to estimate community-level effects of HIV prevention and treatment.

A second approach to transmission tree reconstruction is to weight possible infector-infectee links using a pseudolikelihood based on genetic and epidemiologic data. Ypma~\emph{et al.}~\cite{Ypma2012} analyzed a 2003 influenza A(H7N7) outbreak among poultry farms in the Netherlands by combining data on the times of infection and culling at each farm, the distances between the farms, and RNA consensus sequences. The weight of each possible transmission link was the product of components based on temporal, geographic, and genetic data. The weight of a complete transmission tree was the product of the edge weights. The \texttt{R} package \texttt{outbreaker} implements an extension of this approach that allows multiple introductions of infection and unobserved cases~\cite{Jombart2014}. Like the spanning tree methods above, these methods model pathogen evolution as a process that occurs at the moment of transmission. Morelli~\emph{et al.}~\cite{Morelli2012} proposed a variation of these methods that allows within-host pathogen evolution by incorporating the times of infection and observation into the likelihood component for the genetic sequence data.

A third approach is to reconstruct the transmission tree by combining a phylogeny with epidemiologic data, which was first done by Cottam \emph{et al.}~\cite{Cottam2006, Cottam2008a} in an investigation of a 2001 foot-and-mouth disease virus (FMDV) outbreak among farms in the United Kingdom (UK). The phylogeny and the transmission tree were linked by considering possible locations of the most recent common ancestors (MRCAs) of viruses sampled from the farms. The probability $p_{ij}$ that farm $i$ infected farm $j$ was calculated using epidemiologic data on the oldest detected FMDV lesion and the dates of sampling and culling on each farm. The weight of each possible transmission network was proportional to the product of the $p_{ij}$ for all edges $i\rightarrow j$. Similar methods were used to track farm-to-farm spread of a 2007 FMDV outbreak~\cite{Cottam2008b}. Gardy~\emph{et al.}~\cite{Gardy2011} combined social network analysis with a phylogeny based on whole-genome sequences to construct a transmission tree for a tuberculosis outbreak in British Columbia. Didelot~\emph{et al.} used the time of the most recent common ancestor (TMRCA) to identify possible person-to-person transmission events in studies of \emph{Clostridium difficile} transmission in the UK~\cite{Didelot2012} and \emph{Helicobacter pylori} transmission in South Africa~\cite{Didelot2013}. In a study of \emph{Mycobacterium tuberculosis} transmission in the Netherlands, Bryant~\emph{et al.}~\cite{Bryant2013} ruled out transmission between individuals whose samples did not share a parent in the phylogeny. 

Recent research has identified problems with using genetic sequence data to reconstruct transmission trees. Simulations by Worby~\emph{et al.}~\cite{Worby2014a, Worby2014b} found that pairwise genetic distances cannot reliably identify sources of infection. Methods based on phylogenies often underestimate the complexity of the relationship between the phylogenetic and transmission trees. Branching events in a phylogeny do not necessarily correspond to transmissions, and the topology of the phylogenetic tree need not be the same as the topology of the transmission tree~\cite{PybusRambaut2009, Ypma2013-Genetics, RomeroSeverson2014}. These differences are especially important for diseases with significant within-host pathogen diversity and long latent or infectious periods~\cite{RomeroSeverson2014, Didelot2014}. 

Ypma~\emph{et al.}~\cite{Ypma2013-Genetics} and Didelot~\emph{et al.}~\cite{Didelot2014} have developed Bayesian methods that enforce consistency between phylogenetic and transmission trees in Markov chain Monte Carlo (MCMC) iterations. More recently, Lau~\emph{et al.}~\cite{LauGibson2015} have outlined a Bayesian integration of epidemiologic and genetic sequence data that uses likelihoods based on survival analysis, but their approach does not use pathogen phylogenies directly, assuming that a single dominant lineage within each host can be transmitted. Here, we build a systematic understanding of the relationship between pathogen phylogenies and transmission trees under much weaker assumptions about within-host evolution, allowing the incorporation of genetic sequence data into frequentist and Bayesian survival analysis of infectious disease transmission data.

\subsection*{Transmission trees and public health}
Reconstruction of transmission trees is most useful to public health if it leads to generalizable scientific insights about disease transmission. The transmission tree from one outbreak does not generalize to future outbreaks, but a phylogeny provides partial information about who-infected-whom. Survival analysis provides a rigorous but flexible statistical framework for infectious disease transmission data that explicitly links parameter estimation to the set of possible transmission trees~\cite{Kenah4, Kenah5, Kenah6}. In this framework, estimates of transmission parameters such as covariate effects on infectiousness and susceptibility and evolution of infectiousness over time in infectious individuals are based on sums or averages over all possible transmission trees. Since a phylogeny linking pathogen samples from infected individuals constrains the set of possible transmission trees, pathogen genetic sequence data can be combined with epidemiologic data to obtain more efficient estimates of transmission parameters. 

\section*{Methods}

\subsection*{General stochastic S(E)IR model}
At any time, each individual $i \in \{1, \ldots, n\}$ is in one of four states: susceptible (S), exposed (E), infectious (I), or removed (R).  Person $i$ moves from S to E at his or her \textit{infection time} $t_i$, with $t_i = \infty$ if $i$ is never infected.  After infection, $i$ has a \textit{latent period} of length $\varepsilon_i$ during which he or she is infected but not infectious.  At time $t_i +\varepsilon_i$, $i$ moves from E to I, beginning an \textit{infectious period} of length $\iota_i$.  At time $t_i+ \varepsilon_i + \iota_i$, $i$ moves from I to R, where he or she can no longer infect others or be infected.  The latent period $\varepsilon_i$ is a nonnegative random variable, the infectious period $\iota_i$ is a strictly positive random variable, and both have finite mean and variance. If person $i$ is infected, the time elapsed since the onset of infectiousness at time $t_i + \varepsilon_i$ is the \emph{infectious age} of $i$. 

After becoming infectious at time $t_i + \varepsilon_i$, person $i$ makes infectious contact with $j \neq i$ at time $t_{ij} = t_i + \varepsilon_i + \tau^*_{ij}$. We define infectious contact to be sufficient to cause infection in a susceptible person, so $t_j \leq t_{ij}$. The \emph{infectious contact interval} $\tau^*_{ij}$ is a strictly positive random variable with $\tau^*_{ij} = \infty$ if infectious contact never occurs.  Since infectious contact must occur while $i$ is infectious or never, $\tau^*_{ij} \in (0, \iota_i]$ or $\tau^*_{ij} = \infty$.  

For each ordered pair $ij$, let $C_{ij} = 1$ if infectious contact from $i$ to $j$ is possible and $C_{ij} = 0$ otherwise. For example, the $C_{ij}$ could be the entries in the adjacency matrix for a contact network. However, we do not require that $C_{ij} = C_{ji}$. We assume the infectious contact interval $\tau^*_{ij}$ is generated in the following way: A \textit{contact interval} $\tau_{ij}$ is drawn from a distribution with hazard function $h_{ij}(\tau)$.  If $\tau_{ij}\leq\iota_i$ and $C_{ij} = 1$, then $\tau^*_{ij} = \tau_{ij}$.  Otherwise, $\tau^*_{ij} = \infty$.  

\paragraph{Epidemiologic data}
Our epidemiologic data contain the times of all $\text{S} \rightarrow \text{E}$ (infection), $\text{E} \rightarrow \text{I}$ (infectiousness onset), and $\text{I} \rightarrow \text{R}$ (removal) transitions in the population between time $0$ and time $T$.  For all ordered pairs $ij$ in which $i$ is infected, we observe $C_{ij}$. 

An \emph{exogenous} infection occurs when an individual is infected from a source outside the observed population. An \emph{endogenous} infection occurs when an individual is infected from within the observed population. For each endogenous infection $j$, let $v_j$ denote the index of his or her infector. Let $v_j = 0$ if $j$ is an exogenous infection and $v_j = \infty$ if $j$ is not infected. Let $\mathcal{V}_j$ denote the set of possible $v_j < \infty$, which we call the \emph{infectious set} of $j$. If $j$ is not infected, let $\mathcal{V}_j = \varnothing$ (the empty set).

The \emph{transmission tree} is the directed network with an edge from $v_j$ to $j$ for each infected $j$.  It is a directed tree rooted at node $0$, and it can be represented by a vector $\mathbf{v} = (v_1,\ldots, v_n)$.  Let $\mathcal{V}$ denote the set of all possible $\mathbf{v}$ consistent with the observed data. A $\mathbf{v}\in\mathcal{V}$ can be generated by choosing a $v_j\in\mathcal{V}_j$ for each infected $j$, but we do not assume that all possible transmission trees have the same probability.

\subsection*{Survival analysis of transmission data}
Survival analysis of infectious disease transmission data can be viewed as a generalization of discrete-time chain binomial models~\cite{RampeyLongini} to continuous time, and it includes parametric methods~\cite{Kenah4}, nonparametric methods~\cite{Kenah5}, and semiparametric relative-risk regression models~\cite{Kenah6}. For simplicity, we use parametric methods and assume that exogenous infections are known. Let the hazard of infectious contact from $i$ to $j$ at time $\tau$ after the onset of infectiousness in $i$ be
\begin{equation}
    h_{ij}(\tau) = \exp\big(\bm{\beta}_0^\top \bm{X}_{ij}(\tau)\big) h_0(\tau),
    \label{eq:endogenous}
\end{equation}
where $\bm{\beta}_0$ is an unknown coefficient vector, $\bm{X}_{ij}(\tau)$ is a covariate vector, and $h_0(\tau)$ is a baseline hazard function. The vector $\bm{X}_{ij}(\tau)$ can include individual-level covariates affecting the infectiousness of $i$ or the susceptibility of $j$ as well as pairwise covariates (e.g., membership in the same household). The coefficient vector $\bm{\beta}_0$ captures covariate effects on the hazard of transmission, and the baseline hazard function $h_0(\tau)$ captures the evolution of infectiousness over time in infectious individuals. 

We assume that $\tau_{ij}$ can be observed only if $j$ is infected by $i$ at time $t_i + \varepsilon_i + \tau_{ij}$. The contact interval $\tau_{ij}$ will be unobserved if $i$ recovers from infectiousness before making infectious contact with $j$, if $j$ is infected by a someone other than $i$, or if observation of $j$ has stopped. Let $I_i(\tau) = \indicator_{\tau \in (0, \iota_i]}$ be a left-continuous process indicating whether $i$ remains infectious at infectious age $\tau$. Let $S_{ij}(\tau) = \indicator_{t_i + \varepsilon_i + \tau \leq t_j}$ be a left-continuous process indicating whether $j$ remains susceptible when $i$ reaches infectious age $\tau$. Assume that the population is under observation until a stopping time $T$ and let $O_{ij}(\tau) = \indicator_{t_i + \varepsilon_i + \tau \leq T}$ be a left-continuous process indicating whether $j$ is under observation when $i$ reaches infectious age $\tau$. Then 
\begin{equation}
    Y_{ij}(\tau) = C_{ij} I_i(\tau) S_{ij}(\tau) O_{ij}(\tau)
\end{equation}
is a left-continuous process indicating whether infectious contact from $i$ to $j$ can be observed at infectious age $\tau$ of $i$. The assumptions above ensure that censoring of $\tau_{ij}$ is independent for all $ij$, and they can be relaxed if independent censoring is preserved.

Let $\bm{\theta}$ be a parameter vector for a family of hazard functions $h(\tau, \bm{\theta})$ such that $h_0(\tau) = h(\tau, \bm{\theta}_0)$ for an unknown $\bm{\theta}_0$. To allow maximum likelihood estimation, we assume that $h(\tau, \bm{\theta})$ has continuous second derivatives with respect to $\bm{\theta}$. Let 
\begin{equation}
    h_{ij}(\tau, \bm{\beta}, \bm{\theta}) = \exp\big(\bm{\beta}^\top \bm{X}_{ij}(\tau)\big) h(\tau, \bm{\theta}).
\end{equation}
Let $\mathcal{W}_j = \{i: t_i + \varepsilon_i < t_j \text{ and } C_{ij} = 1\}$ denote the set of all infectious individuals to whom $j$ was exposed while susceptible, which we call the \emph{exposure set} of $j$. When we observe who-infected-whom (i.e., $\mathbf{v}$ is known), the likelihood is
\begin{equation}
    L_\mathbf{v}(\bm{\beta}, \bm{\theta}) = \prod_{j = 1}^n \bigg[h_{v_j j}(t_j - t_{v_j} - \varepsilon_{v_j}, \bm{\beta}, \bm{\theta})^{\indicator_{v_j \not\in \{0, \infty\}}} \prod_{i \in \mathcal{W}_j} e^{-\int_0^{\iota_i} h_{ij}(\tau, \bm{\beta}, \bm{\theta}) Y_{ij}(\tau) \dif \tau}\bigg].
    \label{eq:Lv}
\end{equation}
The hazard terms depend on $\mathbf{v}$, but the survival terms do not~\cite{Kenah4}. 

When we do not observe who-infected-whom, the likelihood is a sum over all possible transmission trees: $L(\bm{\beta}, \bm{\theta}) = \sum_{\mathbf{v} \in \mathcal{V}} L_\mathbf{v}(\bm{\beta}, \bm{\theta})$~\cite{Kenah4}. Each $\mathbf{v} \in \mathcal{V}$ can be generated by choosing a $v_j \in \mathcal{V}_j$ for each endogenous infection $j$. Given the epidemiologic data, each $v_j$ can be chosen independently~\cite{Kenah3}. This leads to the sum-product factorization
\begin{equation}
    L(\bm{\beta}, \bm{\theta}) = \prod_{j = 1}^n \bigg[\Big(\sum_{i \in \mathcal{V}_j} h_{ij}(t_j - t_i - \varepsilon_i, \bm{\beta}, \bm{\theta})\Big)^{\indicator_{v_j \not\in \{0, \infty\}}} \prod_{i \in \mathcal{W}_j} e^{-\int_0^{\iota_i} h_{ij}(\tau, \bm{\beta}, \bm{\theta}) Y_{ij}(\tau) \dif \tau}\bigg].
    \label{eq:L} 
\end{equation}
The probability of a particular transmission tree $\mathbf{v}$ is
\begin{equation}
    \Pr(\mathbf{v} | \bm{\beta}, \bm{\theta}) = \frac{L_\mathbf{v}(\bm{\beta}, \bm{\theta})}{L(\bm{\beta}, \bm{\theta})} = \prod_{j: v_j \not\in \{0, \infty\}} \frac{h_{v_j j}(t_j - t_{v_j} - \varepsilon_{v_j}, \bm{\beta}, \bm{\theta})}{\sum_{i \in \mathcal{V}_j} h_{ij}(t_j - t_i - \varepsilon_i, \bm{\beta}, \bm{\theta})},
\end{equation}
and $L_\mathbf{v}(\bm{\beta}, \bm{\theta}) = \Pr(\mathbf{v} | \bm{\beta}, \bm{\theta}) L(\bm{\beta}, \bm{\theta})$. In this framework, estimation of $(\bm{\beta}$, $\bm{\theta})$, and the probabilities of possible transmission trees is simultaneous. An interesting special case is when $h_{ij}(\tau, \bm{\beta}, \bm{\theta}) = \lambda$ for all $ij$. Then $\Pr(\mathbf{v} | \bm{\beta}, \bm{\theta})$ does not depend on $\lambda$, so the transmission tree is an ancillary statistic~\cite{Kenah4}.

\paragraph{Likelihood calculation with a phylogeny}
Let $\Phi$ denote a pathogen phylogeny, $\mathbf{v}$ denote a transmission tree, and $Epi$ denote the epidemiologic data. Let the function $\Pr(\cdot)$ denote probabilities or probability densities as necessary, and let $\mathcal{V}$ denote the set of transmission trees consistent with both $\Phi$ and $Epi$. Then the likelihood for $\bm{\beta}$ is
\begin{equation}
    \Pr(\Phi, Epi | \bm{\beta}, \bm{\theta}) 
    = \sum_{\mathbf{v} \in \mathcal{V}} \Pr(\mathbf{v}, \Phi, Epi | \bm{\beta}, \bm{\theta})
    = \sum_{\mathbf{v} \in \mathcal{V}} \Pr(\Phi | \mathbf{v}, Epi) \Pr(\mathbf{v}, Epi | \bm{\beta}, \bm{\theta}).
    \label{eq:Lphy}
\end{equation}
The factor $\Pr(\mathbf{v}, Epi | \bm{\beta}, \bm{\theta})$ is the likelihood in equation~\eqref{eq:Lv}. The factor $\Pr(\Phi | \mathbf{v}, Epi)$ depends on within-host pathogen evolution and could incorporate genetic distances and branching times, allowing the joint estimation of between-host transmission parameters and within-host evolutionary parameters. For example, within-host coalescent models were used by Ypma~\emph{et al.}~\cite{Ypma2013-Genetics} and Didelot~\emph{et al.}~\cite{Didelot2014}. The probability of a given transmission tree $\mathbf{v}$ is
\begin{equation}
    \Pr(\mathbf{v} | \bm{\beta}, \bm{\theta}, \Phi) = \frac{\Pr(\Phi | \mathbf{v}, Epi) \Pr(\mathbf{v}, Epi | \bm{\beta}, \bm{\theta})}{\Pr(\Phi, Epi | \bm{\beta}, \bm{\theta})}.
\end{equation}
As before, estimation of $(\bm{\beta}, \bm{\theta})$ and the probabilities of different transmission trees is simultaneous. By providing partial information about who-infected-whom, a phylogeny can increase the precision of transmission parameter estimates.

\subsection*{Phylogenies and transmission trees}
The relationship between phylogenies and transmission trees we develop here is similar to the approach taken by Cottam~\emph{et al.}~\cite{Cottam2008a} who linked phylogenetic and transmission trees via the locations of common ancestors. It is logically equivalent to the approaches of Ypma~\emph{et al.}~\cite{Ypma2013-Genetics} who joined the within-host phylogenies of infectors and infectees into a single phylogeny, Didelot~\emph{et al.}~\cite{Didelot2014} who colored lineages in the phylogeny with a unique color for each individual, and Hall and Rambaut~\cite{HallRambaut2015} who represented transmission trees as partitions of phylogenies. We begin with these assumptions:
\begin{enumerate}   
    \item \label{assum:once} Each individual is infected at most once.
    \item \label{assum:single} Each infection is initiated by a single pathogen. Following infection, within-host pathogen evolution occurs and the evolved pathogens are transmitted to others.
    \item \label{assum:order} The order in which infections (or onsets of infectiousness) occurred is known.
    \item \label{assum:tree} We have at least one pathogen sequence from each infected individual, and these sequences are linked in a rooted phylogeny. The root of this phylogeny has a parent node $r_0$.
    \item \label{assum:trans} Each node in the phylogeny represents a pathogen that had a host, which is also the ``host'' of the node. A parent-child relationship between nodes with different hosts represents a direct transmission of infection from the host of the parent to the host of the child. The node $r_0$ has a host outside the observed population.
\end{enumerate}
The first two assumptions concern the biology of disease. The last three assumptions concern the resolution of the epidemiologic data, which can be controlled through study design. Initially, we use only the topology of the pathogen phylogeny to infer the set of possible transmission trees. Later, we consider how branching times at interior nodes further restrict the set of possible transmission trees.

\paragraph{Transmission trees and interior node hosts}
The leaves (tips) of the phylogenetic tree represent sampled pathogens. Each interior node represents a most recent common ancestor (MRCA) of two or more sampled pathogens. Let $\host(x)$ be the host of the pathogen represented by node $x$ in the phylogeny. If $x$ is a leaf, then $\host(x)$ is known. If $x$ had a host outside the observed population, let $\host(x) = 0$. In particular, $\host(r_0) = 0$. The hosts of all other interior nodes are unknown. 

\begin{lemma}
    The nodes hosted by an infected individual form a subtree of the phylogenetic tree. 
    \label{lem:subtree}
\end{lemma}
\begin{proof}
    See~\nameref{S1app}.
\end{proof}

\begin{theorem}
    A phylogeny with known interior node hosts implies a unique transmission tree. 
    \label{thm:surjection}
\end{theorem}
\begin{proof}
    See~\nameref{S1app}.
\end{proof}

Lemma~\ref{lem:subtree} applies to nodes hosted by infected individuals in the observed population, not to the set of nodes hosted by $0$ (i.e., hosts outside the observed population). In the rest of this section, we assume that the set of nodes hosted by $0$ is a tree rooted at $r_0$. In practice, this restricts the study design. For example, this assumption would be violated if we observed only individuals $A$ and $C$ in a household where $A$ was the index case and there was a chain of transmission $A \rightarrow B \rightarrow C$. The results of this section can be generalized to phylogenies where the set of nodes hosted by $0$ is a union of subtrees as long as each subtree has a known root node. In this case, the phylogeny can split into disjoint pieces by erasing the incoming edge to each root of a subtree hosted by $0$. Each of these pieces can be treated as a separate phylogeny in which the nodes hosted by $0$ form a subtree. 

An assignment of interior node hosts consistent with Lemma~\ref{lem:subtree} will produce at most one transmission tree. A \emph{possible assignment of interior node hosts} is an assignment consistent with Lemma~\ref{lem:subtree} that produces a transmission tree consistent with the epidemiologic data. A \emph{possible transmission tree} is a transmission tree consistent with the epidemiologic data that can be produced by at least one assignment of interior node hosts consistent with Lemma~\ref{lem:subtree}. We now show that each possible transmission tree is produced by exactly one possible assignment of interior node hosts. 

Let $C_x$ denote the set of nodes in the phylogenetic clade rooted at node $x$, and let $L_x$ be the set of hosts of leaf nodes in $C_x$. If $x$ is an interior node, $\host(x)$ may not be in $L_x$. Let $\first(x)$ denote the individual in $L_x$ who is infected earliest or has the earliest onset of infectiousness, at least one of which is well-defined by Assumption~\ref{assum:order}. If the individual infected earliest and the individual with the earliest onset of infectiousness are different, either of them can be used as $\first(x)$. If $x$ is a leaf, $C_x = \{x\}$, $L(x) = \{\host(x)\}$, and $\first(x) = \host(x)$.

\begin{lemma}
    For any node $x$, $\host(x) = \first(x)$ or $\host(x)$ infected $\first(x)$.
    \label{lem:first}
\end{lemma}
\begin{proof}
    See~\nameref{S1app}.
\end{proof}

\begin{theorem}
    A transmission tree corresponds to at most one possible assignment of interior node hosts in a phylogeny.
    \label{thm:injection}
\end{theorem}
\begin{proof}
    See~\nameref{S1app}.
\end{proof}

\begin{figure}
    \centering
    \resizebox{.9\textwidth}{!}{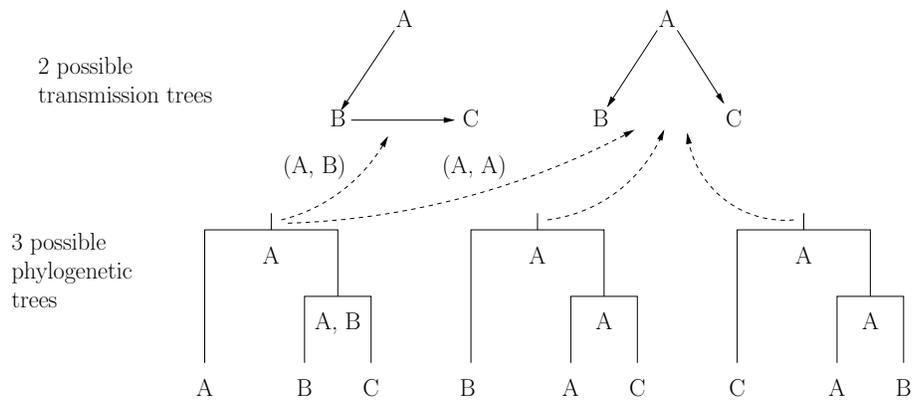}
    \caption{{\bf Two possible transmission trees and three possible pathogen phylogenies for a household outbreak.} A, B, and C were infected in alphabetical order such that their infectious sets are $\mathcal{V}_A = \{0\}$, $\mathcal{V}_B = \{A\}$, and $\mathcal{V}_C = \{A, B\}$. We have a single pathogen sequence from each person. The top shows the two possible transmission trees within the household: either A infected B and B infected C (left) or A infected B and C (right). The bottom shows the three rooted, bifurcating phylogenies linking pathogen sequences from A, B, and C. In each phylogeny, the possible hosts are written underneath each interior node and arrows indicate how each assignment of interior node hosts determines a transmission tree via Assumption~\ref{assum:trans}.}
    \label{fig:ABC}
\end{figure}

Theorems~\ref{thm:surjection} and~\ref{thm:injection} imply a one-to-one correspondence between the possible transmission trees and the possible assignments of interior node hosts in a phylogeny. Fig~\ref{fig:ABC} illustrates this relationship in a very simple case. An similar result was proven independently by Hall and Rambaut~\cite{HallRambaut2015} using partitions of phylogenies.

\paragraph*{Sets of possible interior node hosts}
Theorems~\ref{thm:surjection} and~\ref{thm:injection} reduce the problem of finding the transmission trees consistent with a given phylogeny to that of finding the possible assignments of interior node hosts. Let $H_x$ denote the set of hosts $h$ of node $x$ such that at least one possible transmission tree can be generated when $\text{host}(x) = h$. There are two sets of constraints on $H_x$. The ancestors of $x$ constrain the possible hosts because of Lemma~\ref{lem:subtree}. The descendants of $x$ constrain the possible hosts because $\host(x)$ must be a common ancestor of all members of $L_x$ in the transmission tree. 

We deal first with the descendant constraints. A \emph{transmission tree within clade $C_x$} is a transmission tree rooted at $\host(x)$ that consists of $\host(x)$ and all members of $L_x$. Let $D_x$ denote the set of all hosts $h$ such that at least one possible transmission tree within $C_x$ can be generated when $\host(x) = h$. If $x$ is a leaf, $D_x = \{\host(x)\}$ where $\host(x)$ is the source of the pathogen sample whose genetic sequence is represented by $x$. If $x$ is an interior node, $D_x$ can be calculated using the following results:

\begin{lemma}
    If $x$ is an interior node, $\host(x) = \first(x)$ or $\host(x) = \host\big(\parent(x)\big)$.
    \label{lem:parent}
\end{lemma}
\begin{proof}
    See~\nameref{S1app}.
\end{proof}
\begin{lemma}
    If $x$ is an interior node with child $y$ in the phylogeny, then 
    \begin{equation}
        \host(x) \in D_y^* = \begin{cases}
            D_y &\text{if } \first(y) \not\in D_y, \\
            D_y \cup \mathcal{V}_{\first(y)} &\text{if } \first(y) \in D_y.
        \end{cases}
        \label{eq:D*}
    \end{equation}
    \label{lem:child}
\end{lemma}
\begin{proof}
    See~\nameref{S1app}.
\end{proof}

\begin{theorem}
For any interior node $x$ in the phylogeny,
\begin{equation}
    D_x = \bigcap_{y \in \text{children}(x)} D_y^*,
    \label{eq:Dx}
\end{equation}
    where $\children(x)$ denotes the children of $x$.
    \label{thm:Dx}
\end{theorem}
\begin{proof}
    See~\nameref{S1app}.
\end{proof}

Now we consider the ancestral constraints on $H_x$. By assumption, $\host(r_0) = 0$. Let $A_{r_0} = \{0\}$. For all other nodes in the phylogeny, let
\begin{equation}
    A_x = H_{\parent(x)} \cup \{\first(x)\}.
\end{equation}

\begin{theorem}
    $H_x = A_x \cap D_x$.
    \label{thm:Hx}
\end{theorem}
\begin{proof}
    See~\nameref{S1app}.
\end{proof}

Since $D_x$ is known for each leaf $x$, Theorem~\ref{thm:Dx} shows that $D_x$ can be found for each interior node $x$ in a postorder (i.e., childen before parents) traversal of the phylogeny. Since $A_{r_0} = \{0\}$, Theorem~\ref{thm:Hx} shows that we can calculate $H_x$ for each interior node $x$ in a preorder (parents before children) traversal of the phylogeny once all $D_x$ are known. Combining these results, we get Algorithm~\ref{alg:host} for calculating all $H_x$ in two traversals of the phylogeny. We say that a phylogeny $\Phi$ is \emph{topologically consistent} with the epidemiologic data if $H_x$ is nonempty for each interior node $x$ of $\Phi$.

\begin{algorithm}
    \normalsize
    \KwIn{Rooted phylogeny $\Phi$ and epidemiologic data}
    \KwOut{$H_x$ for each node $x$ of $\Phi$}
    \For{node $x$ in postorder traversal of $\Phi$}{
        \lIf{$x$ is a leaf}{$D_x = \{\host(x)\}$}
        \lElse{$D_x = \cap_{y \in \children(x)} D^*_y$, where $D^*_y$ is defined in equation~\eqref{eq:D*}}
    }
    \For{node $x$ in preorder traversal of $\Phi$}{
        \lIf{$x = r_0$}{$H_x = \{0\}$}
        \lElse{$H_x = D_x \cap A_x$, where $A_x = H_{\parent(x)} \cup \{\first(x)\}$}
    }
    \caption{{\bf Finding host sets.}}
    \label{alg:host}
\end{algorithm}

Having found the $H_x$ at each interior node $x$, it is possible to generate all possible transmission trees consistent with the phylogeny and the epidemiologic data. 

\begin{theorem}
Given a pathogen phylogeny $\Phi$ that is topologically consistent with the epidemiologic data, a transmission tree $\mathbf{v}$ is possible if and only if it can be generated using Algorithm~\ref{alg:tree}.
    \label{thm:trees}
\end{theorem}
\begin{proof}
    See~\nameref{S1app}.
\end{proof}

\begin{algorithm}
    \normalsize
    \KwIn{Rooted phylogeny $\Phi$ with nonempty $H_x$ for each node $x$}
    \KwOut{Transmission tree $\mathbf{v}$ consistent with $\Phi$ and the epidemiologic data}
    \For{node $x$ in preorder traversal of $\Phi$}{
        \lIf{$x = r_0$}{set $\host(x) = 0$}
        \Else{
            $w = \parent(x)$\;
            choose $\host(x) \in H_x \cap \{\host(w), \first(x)\}$\;
            \If{$\host(x) \neq \host(w)$}{
                add edge $\host(w) \rightarrow \host(x)$ to $\mathbf{v}$, adding node $\host(x)$ if necessary
            }
        }
    }
    \caption{{\bf Generating transmission trees.}}
    \label{alg:tree}
\end{algorithm}

Theorem~\ref{thm:trees} gives a useful indication of the value of a phylogeny. A bifurcating phylogeny with $n$ leaves has $n - 1$ interior nodes. For each interior node $x$, we have at most $2$ possible hosts given the host of $\parent(x)$. Thus, there are at most $2^{n - 1}$ possible transmission trees consistent with a pathogen phylogeny of $n$ infections. Without a phylogeny, the worst-case scenario is $n!$ possible transmission trees. Using partitions, Hall and Rambaut~\cite{HallRambaut2015} independently proved that a phylogeny reduces the number of possible transmission trees when $n > 2$.

The combination of Algorithms~\ref{alg:host} and~\ref{alg:tree} is similar to a Sankoff parsimony~\cite{Sankoff1975} where the states represent infected individuals and the cost of going from state $i$ to state $j$ is $1$ if $i \in \mathcal{V}_j$ and $\infty$ otherwise. If there is a single exogenous infection, any transmission tree consistent with the phylogeny and the epidemiologic data will have cost $n - 1$, and all other transmission trees will have infinite cost. Compared to the more general context of a Sankoff parsimony, our algorithms gain efficiency by not having to consider all possible states at each internal node or calculate costs. They also have the advantage of being based on explicit assumptions about the biology of infection and study design. 

\paragraph{Host sets under a molecular clock}
Under a strict or relaxed molecular clock model, each interior node of the phylogenetic tree can be assigned a branching time based on its genetic distance to one or more sequences with known sampling times. These branching times produce new opportunities for inconsistency between the phylogenetic tree and the epidemic data, increasing the possible value of the phylogeny for transmission parameter estimation. Let $t_x$ denote time assigned to node $x$ in the phylogeny. For each leaf, $t_x$ is the time at which the corresponding pathogen was sampled. If $x$ is an interior node, $t_x$ is a branching time. 

If $\host(x)$ is known, the branching time $t_x$ is subject to two constraints. Because branching must occur after the infection of $\host(x)$ and before the end of his or her infectious period,
\begin{equation}
    t_x \in (t_{\host(x)}, t_{\host(x)} + \varepsilon_{\host(x)} + \iota_{\host(x)}],
\end{equation}
where a square bracket indicates an endpoint included in the interval and a parenthesis indicates an endpoint excluded from the interval. We have $t_x > t_{\host(x)}$ because a branching time in $\host(x)$ cannot occur until after he or she is infected. We have $t_x \leq t_{\text{host}(x)} + \varepsilon_{\text{host}(x)} + \iota_{\text{host}(x)}$ because a descendant of virus $x$ was transmitted to at least one other host. The second constraint is that if $x$ is a parent of $y$ and $\text{host}(x) \neq \text{host}(y)$, then $t_x < t_\text{host}(y)$ because a descendant of virus $x$ infected $\text{host}(y)$. These constraints are sufficient to find a valid set of branching times given a possible assignment of interior node hosts. 

\begin{theorem}
    If a transmission tree is generated using Algorithm~\ref{alg:tree}, then Algorithm~\ref{alg:branching} assigns a valid branching time to each internal node of the phylogeny. Any assignment of branching times consistent with the epidemiologic data can be generated this way.
    \label{thm:branching}
\end{theorem}
\begin{proof}
    See~\nameref{S1app}.
\end{proof}

\begin{algorithm}
    \normalsize
    \KwIn{Rooted phylogeny $\Phi$ with known $\host(x)$ for each node $x$}
    \KwOut{Branching time $t_x$ for each node $x$}
    \For{node $x$ in postorder traversal of $\Phi$}{
        \lIf{$x$ is a leaf}{
            set $t_x$ to be the time pathogen $x$ was sampled
        }
        \Else{
            $t_\text{max} = \min_{y \in \children(x)} t_y$\;        
            choose $t_x \in (t_{\host(x)}, t_\text{max})$\;
        }
    }
    \caption{Assigning branching times.}
    \label{alg:branching}
\end{algorithm}

Now suppose we have a phylogenetic tree with known branching times but unknown interior node hosts. Let 
\begin{equation}
    H(t) = \{i : t_i < t \leq t_i + \varepsilon_i + \iota_i\}
\end{equation}
be the set of individuals who are infected but not yet removed at time $t$, and let $H_x(t)$ be the set of possible hosts of node $x$ when $t_x = t$. To find $H_x(t_x)$, it is not sufficient to calculate $H_x$ using Algorithm~\ref{alg:host} and then let $H_x(t_x) = H_x \cap H(t_x)$. To see why, suppose $x$ and $y$ are nodes in the phylogeny such that $h \in H_x \cap H_y$ and $h \in H(t_x) \cap H(t_y)$. Let $w$ be the MCRA of $x$ and $y$. There is no guarantee that $h \in H_w(t_w)$. If not, $h$ can be in at most one of $H_x(t_x)$ and $H_y(t_y)$ by Lemma~\ref{lem:subtree}.

Algorithm~\ref{alg:host} can be modified to find $H_x(t_x)$ with the following changes: In the postorder traversal, we replace $D_x$ with
\begin{equation}
    D_x(t_x) = H(t_x) \cap \Big(\cap_{y \in \children(x)} D^*_y\Big).
    \label{eq:branchingD}
\end{equation}
In the preorder traversal, we define
\begin{equation}
    A_x = H_{\parent(x)}(t_{\parent(x)}) \cup \{\first(x)\}
\end{equation}
for all $x \neq r_0$. Then $H_x(t_x) = A_x \cap D_x(t_x)$. With these changes, the proofs of Theorems~\ref{thm:Dx} and~\ref{thm:Hx} in~\nameref{S1app} work as before under the additional constraint that $\host(x) \in H(t_x)$ for all nodes $x$ in the phylogeny.

\subsection*{Simulations}
To study the impact of a phylogeny on the efficiency of transmission parameter estimates, we conducted a series of 1,000 simulations. In each simulation, there were 100 independent households of size 6. Each household had an index case with an infection time chosen from an exponential distribution with mean one. Each individual $i$ had a binary covariate $X_i$ that could affect infectiousness and susceptibility. Given a parameter vector $\bm{\beta} = (\beta_\text{inf}, \beta_\text{sus})$, the hazard of infectious contact from $i$ to $j$ at infectious age $\tau$ of $i$ is
\begin{equation}
    h_{ij}(\tau, \bm{\beta}) = \exp(\beta_\text{inf} X_i + \beta_\text{sus} X_j)\lambda_0.
\end{equation}
In each simulation, $\beta_\text{inf}$ and $\beta_\text{sus}$ were independently chosen from a uniform distribution on $(-1, 1)$. In all simulations, the baseline hazard was $\lambda_0 = 1$ and the infectious periods were independent exponential random variables with mean one. 

In each simulation, we analyzed data from the first 200 infections in three ways: using only epidemiologic data via the likelihood in equation~\eqref{eq:L}, using epidemiologic data with who-infected-whom via the likelihood in equation~\eqref{eq:Lv}, and using epidemiologic data with a phylogeny via the likelihood in equation~\eqref{eq:Lphy}. In the phylogenetic analysis, we assumed a single pathogen sample from each infected individual. The within-host phylogeny for each individual who infected $m > 0$ individuals was chosen uniformly at random from all rooted, bifurcating phylogenies with $m + 1$ tips. Within-individual phylogenies were chosen independently and combined into a single phylogeny as in Ypma~\emph{et al.}~\cite{Ypma2013-Genetics}. Thus, the conditional probability $\Pr(\Phi | \mathbf{v}, Epi)$ for a phylogeny $\Phi$ given a transmission tree $\mathbf{v}$ in which each individual $j$ infected $m_j \geq 0$ other individuals was proportional to
\begin{equation}
     \prod_{j: m_j > 0} \frac{2^{m_j - 1} (m_j - 1)!}{(2m_j - 1)!}.
     \label{eq:vweight}
\end{equation}
The set of transmission trees consistent with the phylogeny was determined using Algorithms~\ref{alg:host} and~\ref{alg:tree}. We calculated the mean error, mean squared error, 95\% confidence interval coverage probability, and relative efficiency of $\beta_\text{inf}$, $\beta_\text{sus}$, and $\ln \lambda_0$ estimates in all three analyses.

The simulations were conducted in Python 2.7 (\texttt{www.python.org}) and analysis was conducted in R 3.2 (\texttt{cran.r-project.org}) via RPy2 2.7 (\texttt{rpy.sourceforge.net}). The Python code is in~\nameref{S1txt}. Parameters, point estimates, and 95\% confidence limits are in~\nameref{S1dat}. R code for the simulation data analysis is in~\nameref{S2txt}.

\paragraph{Data on individuals who escape infection} The likelihoods in equations~\eqref{eq:Lv} and~\eqref{eq:L} require data on individuals who were at risk of infection but not infected. Except for Lau~\emph{et al.}~\cite{LauGibson2015}, these data are not used in any of the studies cited in the Introduction. To study the value of data on individuals who escape infection in the households of infected individuals, we repeated all analyses excluding data on the uninfected. The parameters, point estimates, and 95\% confidence limits are in~\nameref{S2dat}. 

\paragraph{Time variation in infectiousness} When contact intervals are exponential and there is no variation in infectiousness, the transmission tree is an ancillary statistic for the hazard $\lambda$ of infectious contact~\cite{Kenah4}. To explore the effect of time variation in infectiousness on the value of a phylogeny, we repeated the simulations using a Weibull contact interval distribution with shape parameter $\gamma = 0.5$. To keep the same mean contact interval, we set the rate parameter $\lambda_0 = 2$. The hazard of infectious contact from $i$ to $j$ at infectious age $\tau$ or $i$ is
\begin{equation}
    h_{ij}(\tau, \bm{\beta}) = \exp(\beta_\text{inf} X_i + \beta_\text{sus} X_j) \gamma \lambda_0^\gamma \tau^{\gamma - 1}.
\end{equation}
With $\gamma = 0.5$, this decreases monotonically during the infectious period. These simulations used the same combinations of $(\beta_\text{inf}, \beta_\text{sus})$ that were used in the simulations with exponential contact intervals. The parameters, point estimates, and 95\% confidence limits are in~\nameref{S3dat}.

\subsection*{Data analysis}
To illustrate an application of these algorithms and likelihoods, we use them to analyze farm-to-farm transmission trees of foot and mouth disease virus (FMDV) in a cluster of 12 epidemiologically linked farms in Durham, UK in 2001. The genetic and epidemiologic data are publicly available as Data S3 and Data S4 in Morelli~\emph{et al.}~\cite{Morelli2012}. These data were previously analyzed by Cottam~\emph{et al.}~\cite{Cottam2008a, Cottam2008b}, Morelli~\emph{et al.}~\cite{Morelli2012}, Ypma~\emph{et al.}~\cite{Ypma2013-Genetics}, and Lau~\emph{et al.}~\cite{LauGibson2015}.

FMDV is a picornavirus that causes a highly contagious disease in cattle, pigs, sheep, and goats~\cite{FMDV2014}. Upon infection, there is an incubation period of approximately 1--12 days in sheep, 2--14 days in cattle, and two or more days in pigs. The incubation period is followed by an acute febrile illness with painful blisters on the feet, the mouth, and the mammary glands. It is transmitted through secretions from infected animals, fomites, virus carried on skin or clothing, and aerosolized virus. Outbreaks of foot-and-mouth disease are difficult to control and can devastate livestock. During the FMDV outbreak, teams from the UK Department for Environment, Food, and Rural Affairs (DEFRA) visited each infected farm~\cite{Cottam2006, Cottam2008a}. They recorded the number and types of susceptible and infected animals, examined infected animals to determine the age of the oldest lesions, and collected epithelial samples. Finally, they recorded the date of the cull. 

\begin{figure}
    \centering
    \includegraphics[width = .75\textwidth]{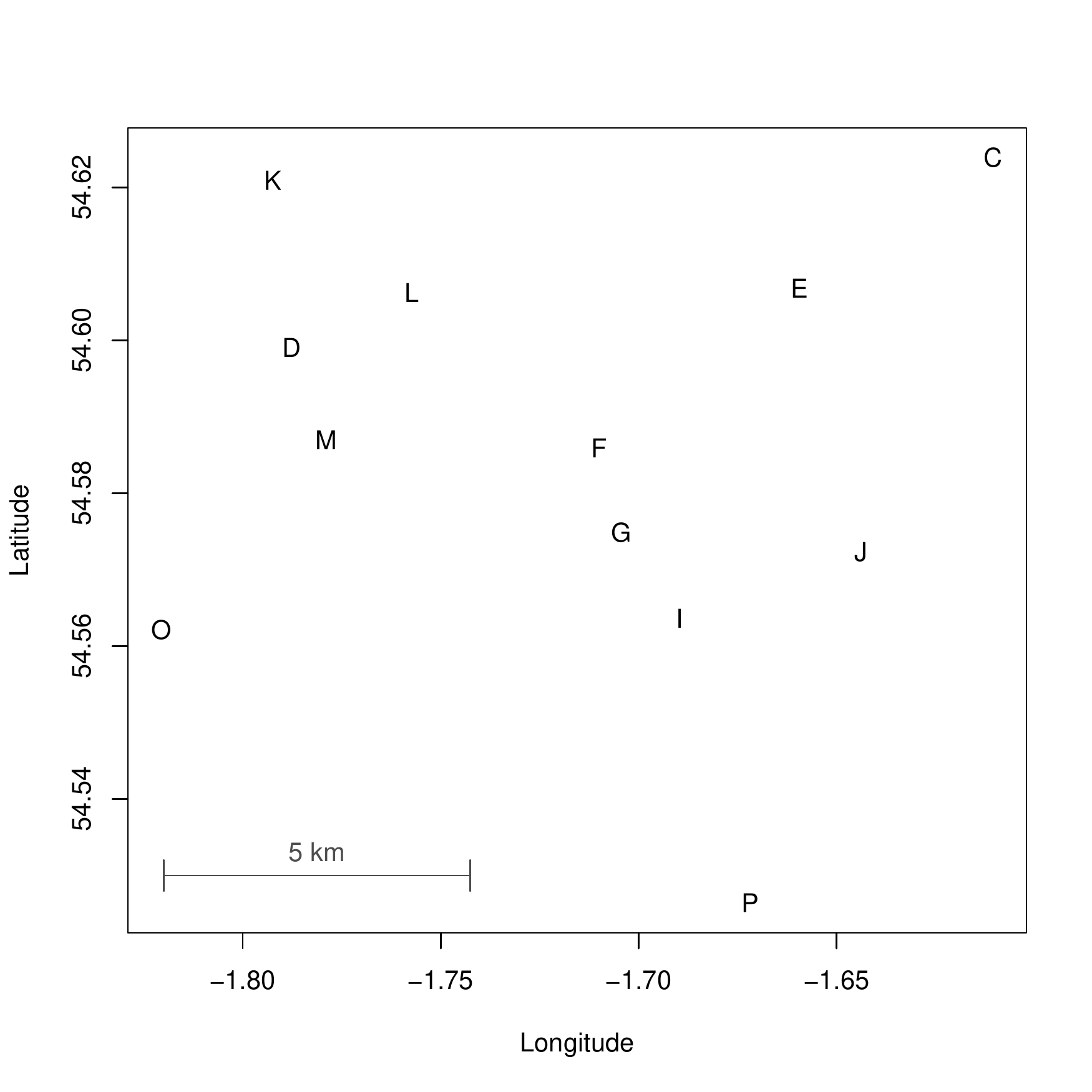}
    \caption{{\bf Relative locations of the 12 farms in the Durham cluster.} These were infected in the 2001 FMDV outbreak in the UK.}
    \label{fig:2001map}
\end{figure}

\begin{figure}
    \centering
    \includegraphics[width = .75\textwidth]{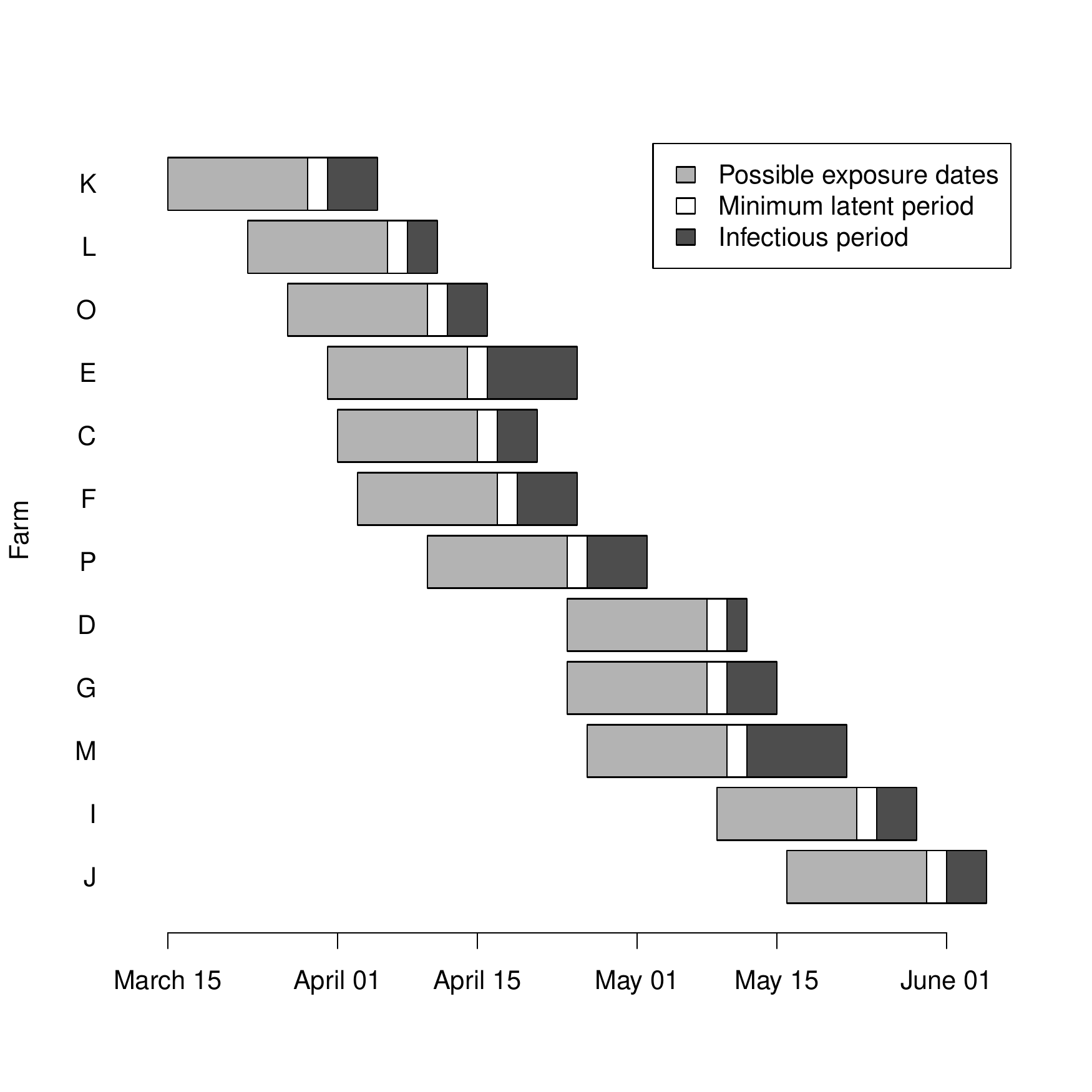}
    \caption{{\bf Timeline of latent and infectious periods in the Durham cluster.} The gray bars represent the range of days on which each farm might have been infected.}
    \label{fig:2001timeline}
\end{figure}

We assume that infectiousness begins on the day that the first lesions appeared and ends with the cull, and we assumed a latent period (between infection and the onset of infectiousness) of 2--16 days. Fig~\ref{fig:2001map} shows the relative locations of the farms, and Fig~\ref{fig:2001timeline} shows the timeline of the latent and infectious periods. Analysis was conducted in R~3.2 (\texttt{cran.r-project.org}), and the code is available in~\nameref{S3txt}.

\paragraph{Estimating the hazard of infectious contact}
Without a phylogeny, we estimated the hazard of transmission from an infected farm to a susceptible farm using the likelihood in equation~\eqref{eq:L}. We assumed a log-logistic contact interval distribution with rate parameter $\lambda$ and shape parameter $\gamma$, which has the hazard function
\begin{equation}
    h(\tau, \lambda, \gamma) = \frac{\gamma \lambda^\gamma \tau^{\gamma - 1}}{1 + (\lambda \tau)^\gamma}
\end{equation}
and the survival function $S(\tau, \lambda, \gamma) = \big(1 + (\lambda \tau)^\gamma\big)^{-1}$ where $\tau$ is the infectious age. This is the simplest parametric model in survival analysis that allows a non-monotonic hazard function. When $\gamma \leq 1$, the hazard decreases monotonically. When $\gamma > 1$, the hazard is unimodal. To enforce the restriction that $\lambda > 0$ and $\gamma > 0$, maximum likelihood estimation was done using parameters $\ln \lambda$ and $\ln \gamma$. 

If farm $j$ is infected on day $t$, let $\mathcal{V}_j(t)$ denote its infectious set and $\mathcal{W}_j(t)$ denote its exposure set. For each $i \in \mathcal{V}_j(t)$, the times $t_i + \varepsilon_i$ of onset of infectiousness and $t_j^*$ of onset of symptoms are known. Without a phylogeny, the overall likelihood contribution for each infected farm $j$ except $K$ (the index farm) is the sum of the likelihood in equation~\eqref{eq:L} over all possible infection times:
\begin{equation}
    \sum_{t = t^*_j - 16}^{t_j^* - 2} \Bigg[\sum_{i \in \mathcal{V}_j(t)} h\big(t - t_i - \varepsilon_i, \lambda, \gamma\big) \prod_{i \in \mathcal{W}_j(t)} S\big(\min(\iota_i, t - t_i - \varepsilon_i), \lambda, \gamma\big)\Bigg].
    \label{eq:Linfected}
\end{equation}
Note that the sum of the hazards is zero for each $t$ where there is no possible infector. Via a sum-product factorization, the product of equation~\eqref{eq:Linfected} over all $j \neq L$ is the sum of the likelihoods of all possible combinations of infection times at the farms. 

With a phylogeny, we estimated the hazard of transmission from an infected farm to a susceptible farm using the likelihood in equation~\eqref{eq:Lphy}, which is a weighted sum of the likelihood contributions of each transmission tree. Let $\mathcal{I}$ denote the set of endogenously infected farms. For each transmission tree $\mathbf{v}$, the likelihood is 
\begin{equation}
    \prod_{j \in \mathcal{I}} \Bigg[\sum_{t = t^*_j - 16}^{t_j^* - 2} h\big(t - t_{v_j} - \varepsilon_{v_j}, \lambda, \gamma\big) \prod_{i \in \mathcal{W}_j(t)} S\big(\min(\iota_i, t - t_i - \varepsilon_i), \lambda, \gamma\big)\Bigg]
\end{equation}
where $v_j$ denotes the infector of farm $j$. We assumed that the within-host phylogeny for each farm that infected $m > 0$ other farms was chosen uniformly at random from all rooted, bifurcating phylogenies with $m + 1$ tips, so each transmission tree gets a weight proportional to equation~\eqref{eq:vweight}. 

\paragraph{Data on farms that escaped infection} As usual, the epidemiologic data set contains only infected farms. To illustrate how our results depend on farms that escaped infection, we repeat the analyses with and without a phylogeny using 6, 12, and 24 uninfected farms. Because observation of the outbreak ended after the end of infectiousness in all infected farms, the likelihood contribution from each uninfected farm is $\prod_{i \in \mathcal{W}_*} S(\iota_i, \lambda, \gamma)$ where $\mathcal{W}_*$ is the set of infected farms and $\iota_i$ is the infectious period for farm $i$.

\section*{Results} 
\subsection*{Simulations}
\begin{figure}
    \centering
    \includegraphics[width = .75\textwidth]{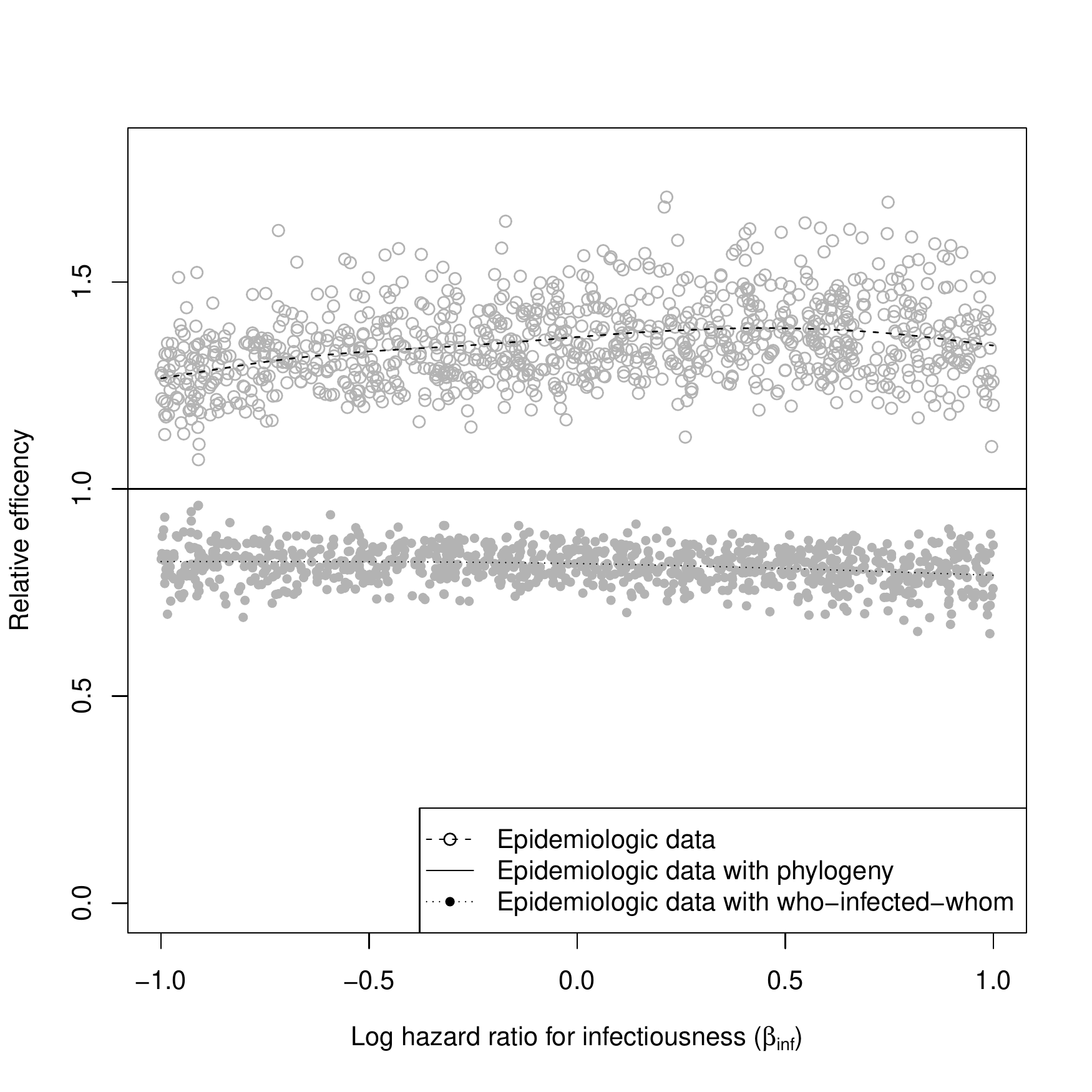}
    \caption{{\bf Relative efficiencies of $\beta_\text{inf}$ estimates based on squared widths of confidence intervals.} The dashed and dotted lines are smoothed means.}
    \label{fig:infRE}
\end{figure}

\begin{figure}
    \centering
    \includegraphics[width = .75\textwidth]{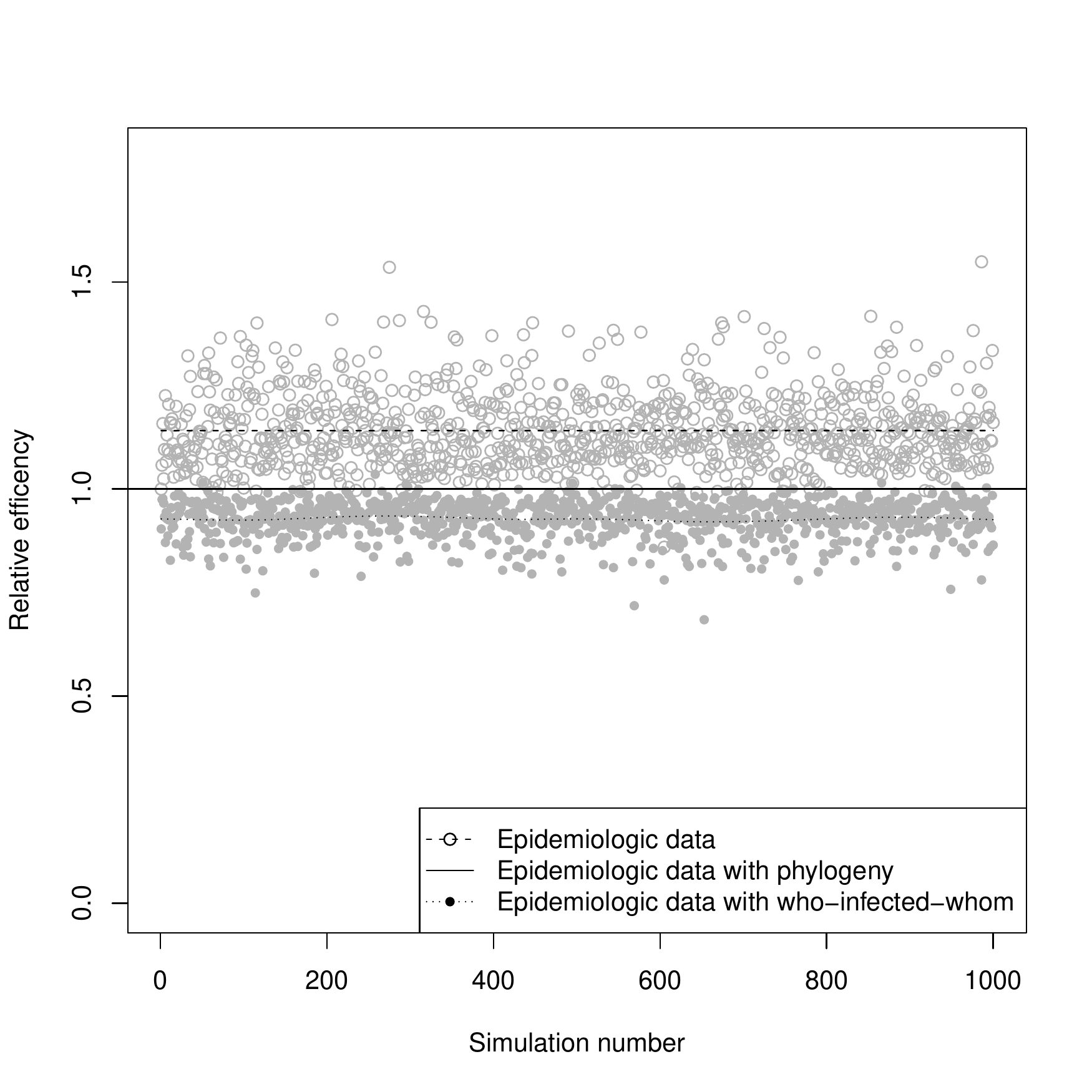}
    \caption{{\bf Relative efficiencies of $\ln \lambda_0$ estimates based on squared widths of confidence intervals.} The dashed and dotted lines are smoothed means.}
    \label{fig:lnlambdaRE}
\end{figure}

Table~\ref{tab:mse} shows the mean error, mean squared error, 95\% confidence interval coverage probability, and relative efficiency of $\beta_\text{inf}$, $\beta_\text{sus}$, and $\ln \lambda_0$ estimators in the simulations. In all cases, the point estimates were nearly unbiased (indicated by the mean error squared being much smaller than the mean squared error) and the 95\% confidence interval coverage probabilities were near $0.95$. Fig~\ref{fig:infRE} shows that estimates of $\beta_\text{inf}$ using a phylogeny were more efficient than estimates using epidemiologic data only and less efficient than estimates using who-infected-whom. By mean squared error, the phylogenetic estimates had a relative efficiency of 1.39 compared to estimates using only epidemiologic data and 0.80 compared to estimates using who-infected-whom. Because knowledge of who-infected-whom does not add to our knowledge of who was infected, all three analyses were equally efficient for $\beta_\text{sus}$ (similar results were obtained for estimates with and without who-infected-whom in Ref~\cite{Kenah6}). Fig~\ref{fig:lnlambdaRE} shows that estimates of $\ln \lambda_0$ using a phylogeny were more efficient than those using epidemiologic data only and less efficient than those using who-infected whom. By mean squared error, the phylogenetic estimates had a relative efficiency of 1.17 compared to estimates using only epidemiologic data and 0.90 compared to estimates using who-infected-whom. 

Table~\ref{tab:mse-infonly} shows the mean error, mean squared error, 95\% confidence interval coverage probability, and relative efficiency of $\beta_\text{inf}$, $\beta_\text{sus}$, and $\ln \lambda_0$ estimators that excluded data on uninfected household members. The mean squared errors were much higher than the corresponding estimators in Table~\ref{tab:mse}, so their relative efficiency was very low. In all cases, the efficiency loss from excluding data on individuals who escape infection was much larger than the efficiency gain from incorporating a phylogeny or from knowing exactly who infected whom. For estimators of $\beta_\text{inf}$ and $\beta_\text{sus}$, the square of the mean error was much smaller than the mean squared error, indicating little bias. Estimates of $\ln \lambda_0$ were biased upward, resulting in extremely low relative efficiencies and coverage probabilities. In Ref~\cite{Kenah4}, similar results were seen for estimates of the basic reproduction number ($R_0$) when approximate likelihoods for mass-action models, which do not require data on uninfected individuals, were used to analyze data from network-based epidemics.

Table~\ref{tab:mse-Weibull} shows results the mean error, mean squared error, 95\% confidence interval coverage probability, and relative efficiency of $\beta_\text{inf}$, $\beta_\text{sus}$, $\ln \lambda_0$, and $\ln \gamma$ estimators from models with Weibull contact interval distributions with rate parameter $\lambda_0 = 2$ and shape parameter $\gamma = .5$. All estimators are unbiased with 95\% confidence interval coverage probabilities near 0.95. The relative efficiencies are similar to those in Table~\ref{tab:mse}, showing that the gains in efficiency for estimates of infectiousness hazard ratios and baseline hazards occur under weak assumptions about the baseline hazard.

\begin{table}
    \caption{Statistical performance of estimators under exponential contact intervals.}
    \begin{tabular}{rccc}
        \hline
        & \textbf{Epidemiologic} & & \\
        & \textbf{only} & \textbf{+ phylogeny} & \textbf{+ who-infected-whom} \\
        \hline
        \multicolumn{4}{l}{$\beta_\text{inf}$} \\
        Mean error & 0.0032 & 0.0060 & 0.0080 \\
        Mean squared error & 0.0537 & 0.0386 & 0.0307 \\
        Coverage probability & 0.945 & 0.942 & 0.945 \\
        Relative efficiency$^*$ & 1 & 1.39 & 1.75 \\
        \hline
        \multicolumn{4}{l}{$\beta_\text{sus}$} \\
        Mean error & 0.0023 & 0.0025 & 0.0023 \\
        Mean squared error & 0.0306 & 0.0306 & 0.0305 \\
        Coverage probability & 0.953 & 0.953 & 0.953 \\
        Relative efficiency$^*$ & 1 & 1.00 & 1.00 \\
        \hline
        \multicolumn{4}{l}{$\ln \lambda_0$} \\
        Mean error & -0.0050 & -0.0054 & -0.0057 \\
        Mean squared error & 0.0300 & 0.0256 & 0.0230 \\
        Coverage probability & 0.952 & 0.953 & 0.955 \\
        Relative efficiency$^*$ & 1 & 1.17 & 1.31 \\
        \hline
    \end{tabular}
    * Compared to estimates with epidemiologic data only.
    \label{tab:mse}
\end{table}

\begin{table}
    \caption{Statistical performance of estimators using infecteds only.}
    \begin{tabular}{rccc}
        \hline
        & \textbf{Epidemiologic} & & \\
        & \textbf{data} & \textbf{+ phylogeny} & \textbf{+ who-infected-whom} \\
        \hline
        \multicolumn{4}{l}{$\beta_\text{inf}$} \\
        Mean error & 0.0300 & 0.0312 & 0.0299 \\
        Mean squared error & 0.1850 & 0.1115 & 0.0804 \\
        Coverage probability & 0.682 & 0.756 & 0.779 \\
        Relative efficiency$^*$ & 0.29 & 0.48 & 0.67 \\
        \hline
        \multicolumn{4}{l}{$\beta_\text{sus}$} \\
        Mean error & 0.0408 & 0.0410 & 0.0261 \\
        Mean squared error & 0.1351 & 0.1339 & 0.1842 \\
        Coverage probability & 0.583 & 0.582 & 0.491 \\
        Relative efficiency$^*$ & 0.23 & 0.23 & 0.17 \\
        \hline
        \multicolumn{4}{l}{$\ln \lambda_0$} \\
        Mean error & 0.7647 & 0.7559 & 0.9341 \\
        Mean squared error & 0.6151 & 0.5978 & 0.9033 \\
        Coverage probability & 0.008 & 0.007 & 0.000 \\
        Relative efficiency$^*$ & 0.05 & 0.05 & 0.03 \\
        \hline
    \end{tabular}
    * Compared to estimates in Table~\ref{tab:mse} with epidemiologic data only.
    \label{tab:mse-infonly}
\end{table}

\begin{table}
    \caption{Statistical performance of estimators under Weibull contact intervals.}
    \begin{tabular}{rccc}
        \hline
        & \textbf{Epidemiologic} & & \\
        & \textbf{data} & \textbf{+ phylogeny} & \textbf{+ who-infected-whom} \\
        \hline
        \multicolumn{4}{l}{$\beta_\text{inf}$} \\
        Mean error & 0.0030 & 0.0043 & 0.0000 \\
        Mean squared error & 0.0454 & 0.0318 & 0.0251 \\
        Coverage probability & 0.948 & 0.960 & 0.966 \\
        Relative efficiency$^*$ & 1 & 1.43 & 1.81 \\
        \hline
        \multicolumn{4}{l}{$\beta_\text{sus}$} \\
        Mean error & -0.0076 & -0.0076 & -0.0075 \\
        Mean squared error & 0.0278 & 0.0276 & 0.0276 \\
        Coverage probability & 0.949 & 0.949 & 0.946 \\
        Relative efficiency$^*$ & 1 & 1.01 & 1.01 \\
        \hline
        \multicolumn{4}{l}{$\ln \lambda_0$} \\
        Mean error & 0.0278 & 0.0224 & 0.0260 \\
        Mean squared error & 0.1188 & 0.1069 & 0.0978 \\
        Coverage probability & 0.941 & 0.948 & 0.943 \\
        Relative efficiency$^*$ & 1 & 1.11 & 1.21 \\
        \hline
        \multicolumn{4}{l}{$\ln \gamma$} \\
        Mean error & 0.0133 & 0.0114 & 0.0106 \\
        Mean squared error & 0.0052 & 0.0046 & 0.0045 \\
        Coverage probability & 0.958 & 0.961 & 0.960 \\
        Relative efficiency$^*$ & 1 & 1.12 & 1.15 \\
        \hline
    \end{tabular}
    * Compared to estimates using epidemiologic data only.
    \label{tab:mse-Weibull}
\end{table}

\subsection*{Data Analysis}
\begin{figure}
    \centering
    \includegraphics[width = .7\textwidth]{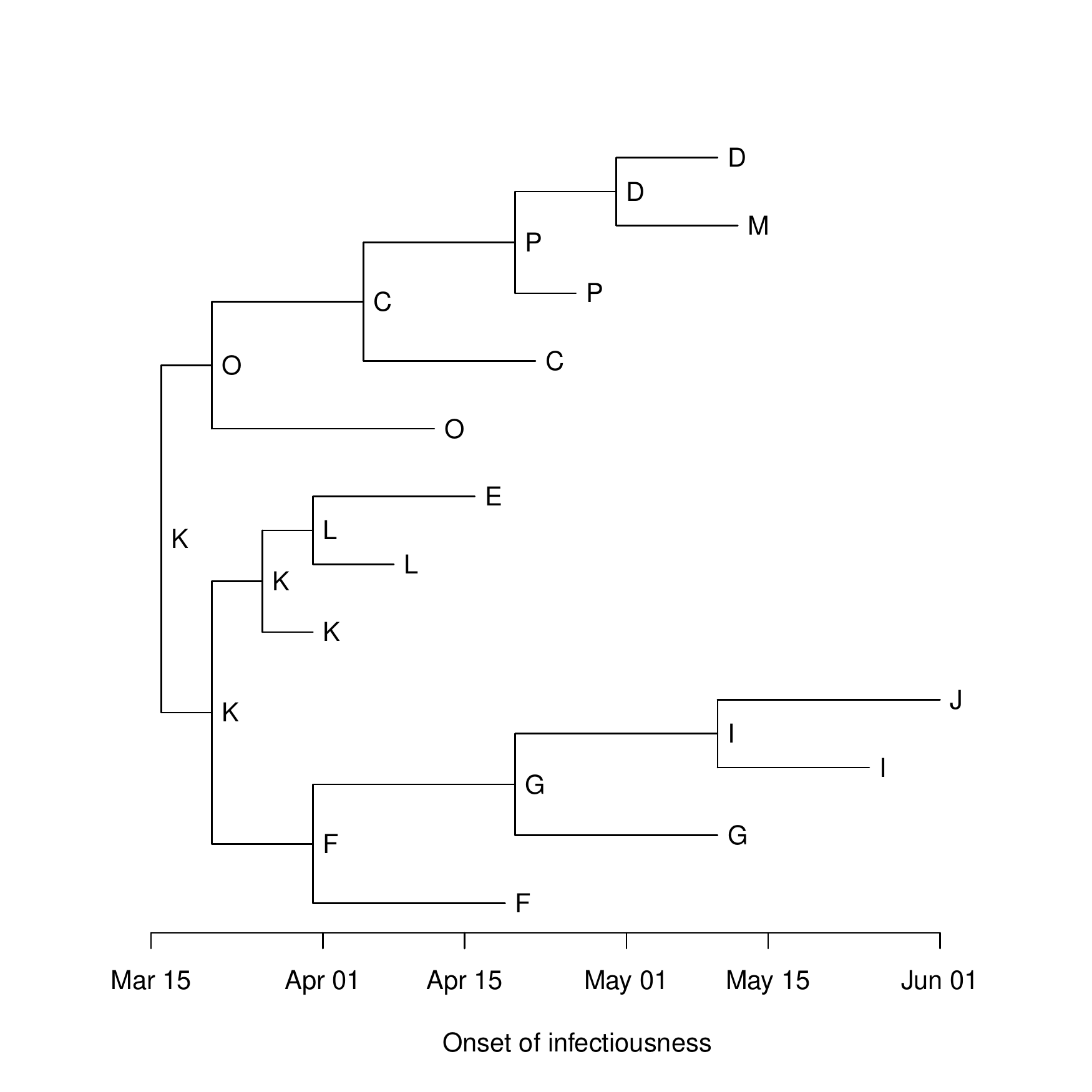}
    \caption{{\bf First hosts in the Durham cluster.} Rooted phylogeny for RNA sequences from the 12 farms in the Durham cluster with tips at the onset of infectiousness. Each interior node $x$ has $\first(x)$ written next to it.}
    \label{fig:2001first}
\end{figure}

\begin{figure}
    \centering
    \includegraphics[width = .7\textwidth]{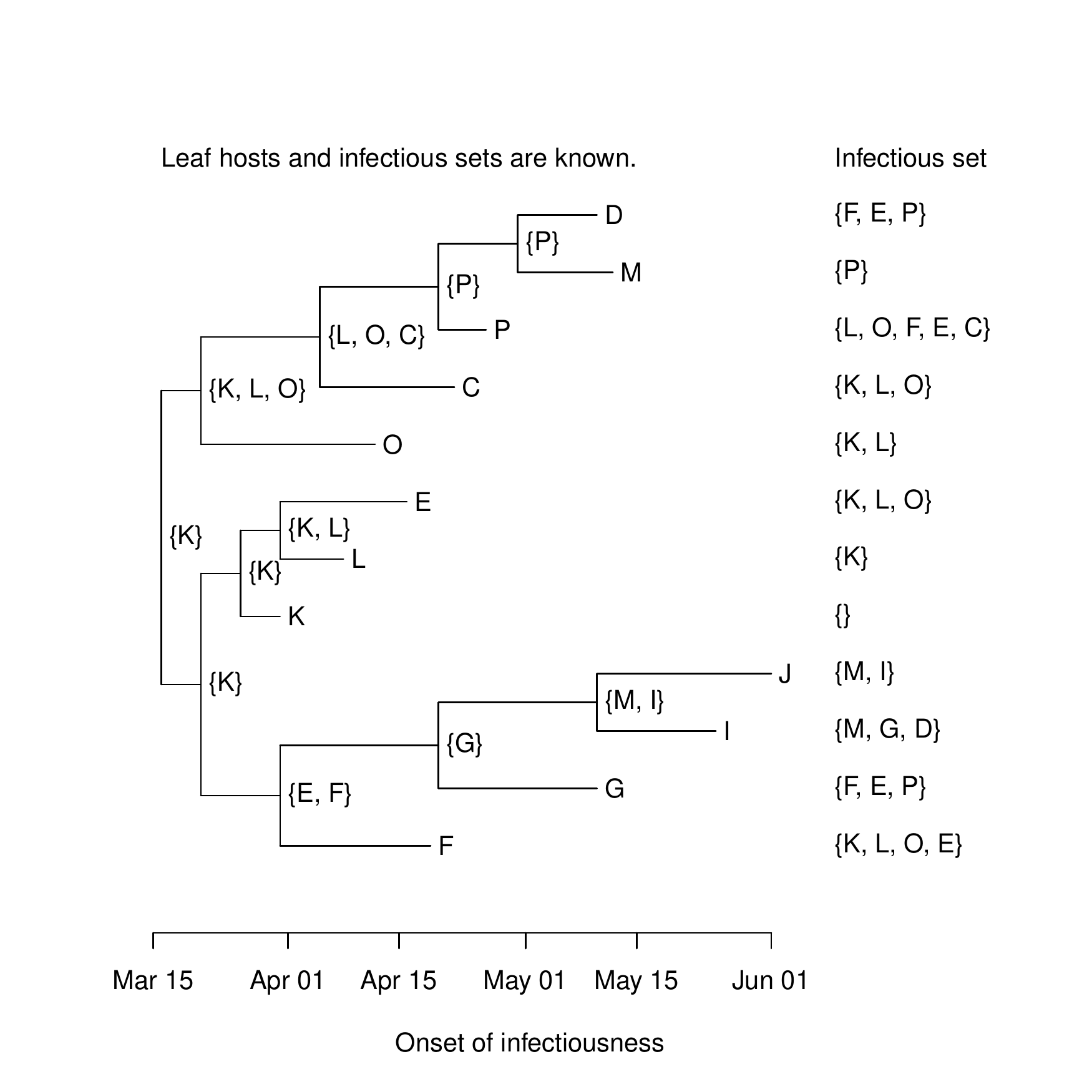}
    \caption{{\bf Postorder host sets in the Durham cluster.} The postorder host set $D_x$ is written next to each interior node $x$. These are calculated in a postorder traversal using the leaf hosts and the infectious sets.}
    \label{fig:2001post}
\end{figure}

\begin{figure}
    \centering
    \includegraphics[width = .7\textwidth]{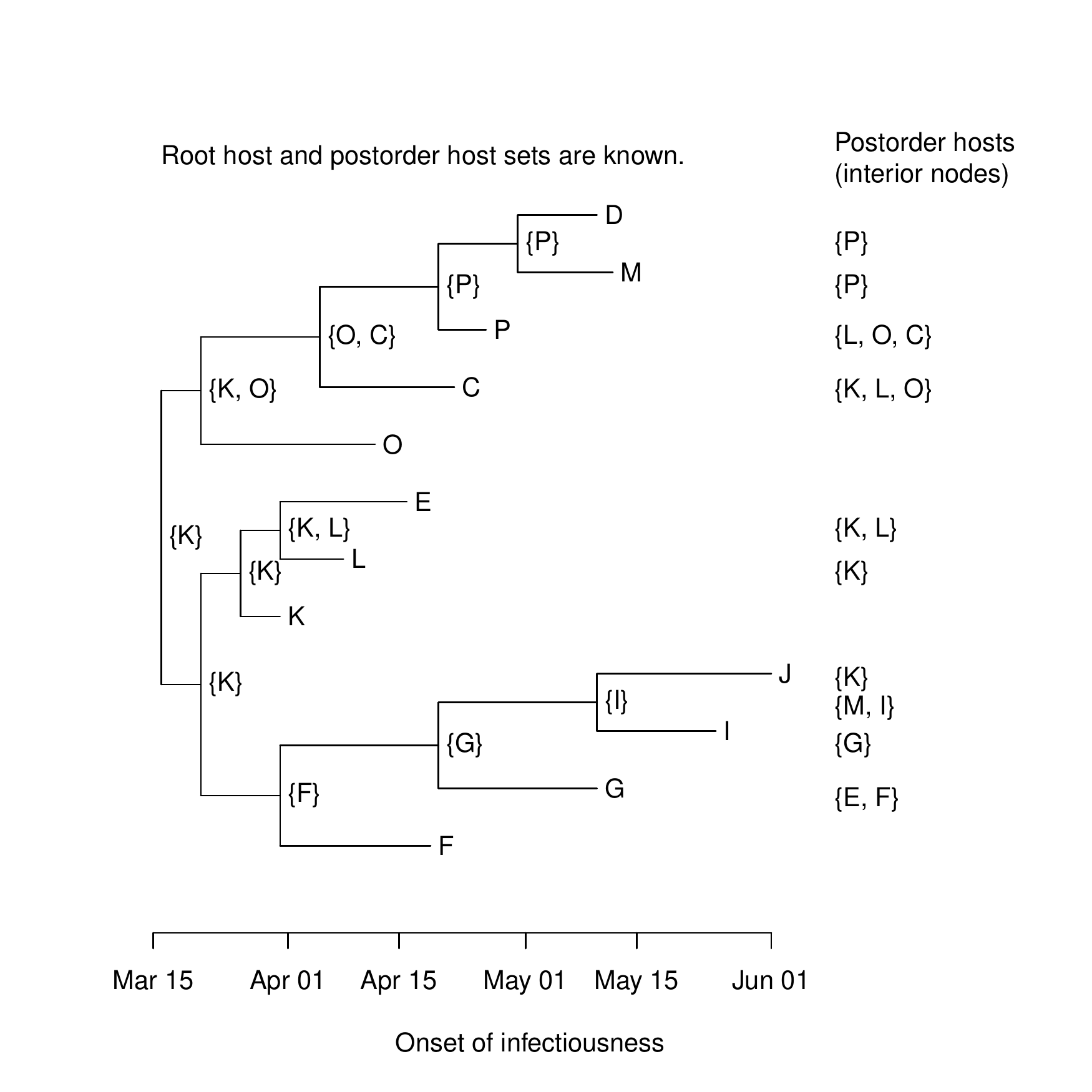}
    \caption{{\bf Host sets in the Durham cluster.} The host set $H_x$ is written next to each interior node $x$. These are calculated in a preorder traversal using the root host and the postorder host sets.}
    \label{fig:2001host}
\end{figure}

\begin{figure}
    \centering
    \includegraphics[width = \textwidth]{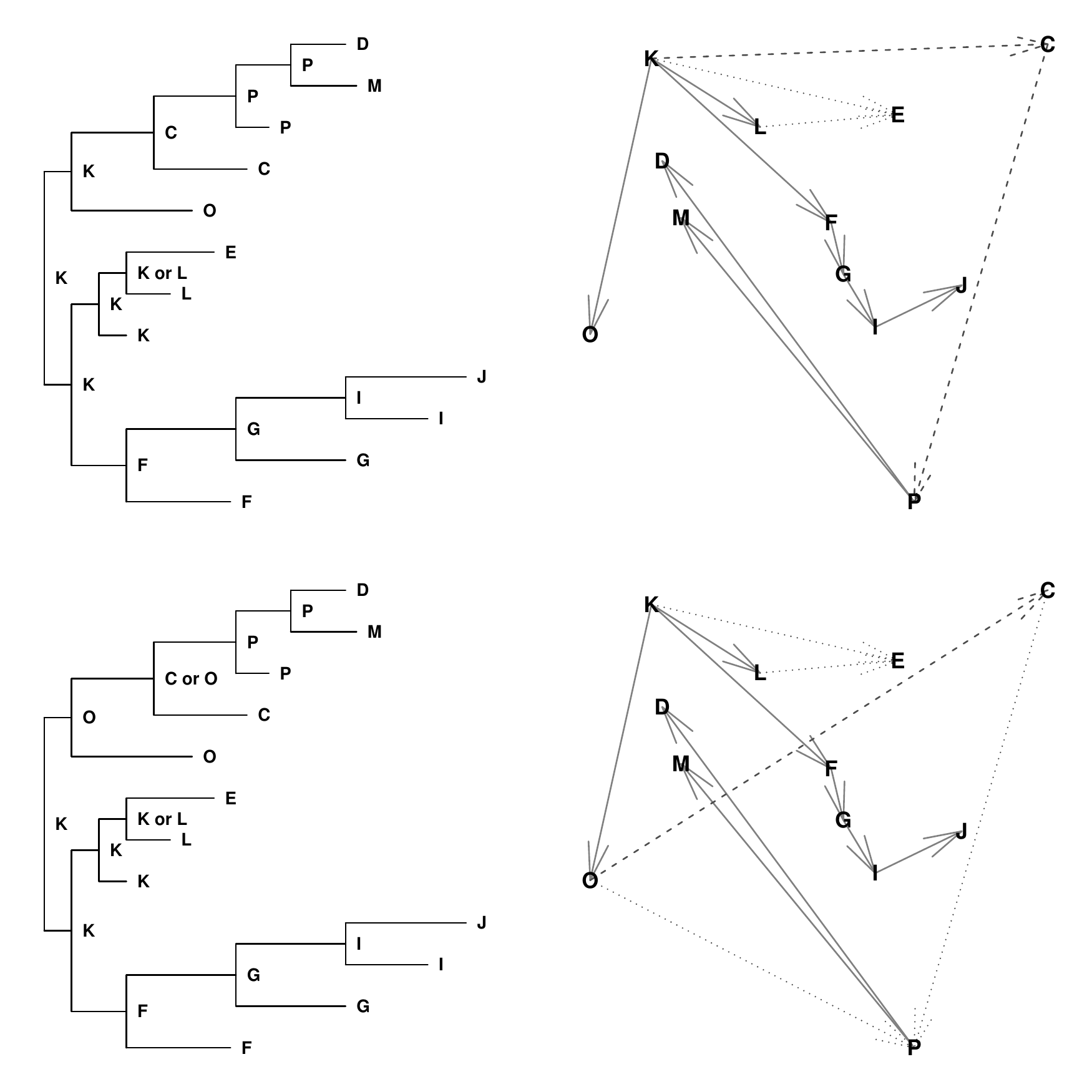}
    \caption{{\bf Interior node host assignments (left) and transmission trees (right) consistent with the phylogeny and the epidemiologic data.} Dashed lines on the right indicate transmissions fixed by the phylogeny to the left. Dotted lines indicate that the infector of an individual depends on a choice of hosts in the phylogeny. At the top, $K$ infects $C$, $C$ infects $P$, and either $K$ or $L$ infects $E$. At the bottom, $O$ infects $C$, either $O$ or $C$ infects $P$, and either $K$ or $L$ infects $E$. There are six possible transmission trees---two on top and four on the bottom.}
    \label{fig:2001trees}
\end{figure}

With no phylogeny, there are $19{,}440$ possible transmission trees linking the 12 farms in the Durham cluster. A phylogeny was constructed in SeaView~\cite{Gouy2010} using consensus RNA sequences from 15 farms, including three farms not epidemiologically linked to the cluster~\cite{Morelli2012}. We used a generalized time reversible (GTR) nucleotide substitution model with four rate classes on $8,196$ sites. Fig~\ref{fig:2001first} shows the rooted phylogeny for the 12 farms in the cluster with branch tips scaled to reflect the time of infectiousness onset at each farm (interior branch lengths do not indicate branching times). The order of infectiousness onsets is known, so $\first(x)$ is the host with the earliest onset of infectiousness in clade $C_x$. Fig~\ref{fig:2001post} shows the postorder host set $D_x$ for each node $x$ in the phylogeny, and Fig~\ref{fig:2001host} shows the host sets. The host is uniquely determined by the phylogeny for all interior nodes except three. Figure~\ref{fig:2001trees} shows the six possible interior node host assignments and the corresponding transmission trees. 

\begin{figure}
    \centering
    \includegraphics[width = .7\textwidth]{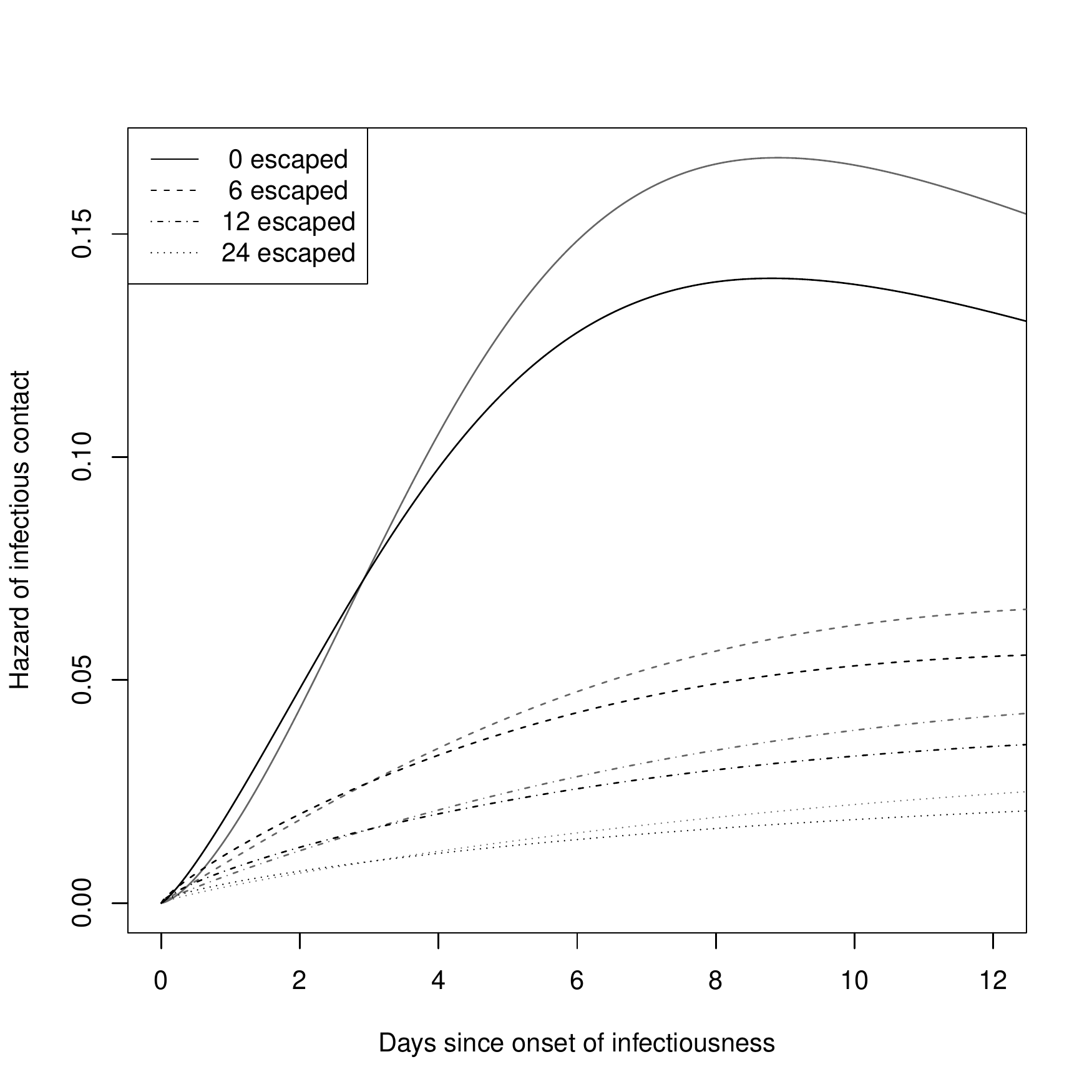}
    \caption{{\bf Predicted farm-to-farm infectiousness with (black) and without (gray) phylogenies.} These are log-logistic hazard functions based on the rate and shape parameters estimates in Table~\ref{tab:hazard}.}
    \label{fig:haz}
\end{figure}

\paragraph{Hazard of infectious contact}
Tab.~\ref{tab:hazard} shows the rate and shape parameter estimates with and without a phylogeny for 0, 6, 12, and 24 uninfected farms. The estimates with a phylogeny have lower rate and shape parameters, suggesting a slightly more rapid increase in infectiousness with a lower peak. The estimates using a phylogeny have narrower confidence intervals. The predicted hazard functions are shown in Fig~\ref{fig:haz}, and they are very sensitive to the number of farms that escaped infection. A similar sensitivity was observed by Lau~\emph{et al.}~\cite{LauGibson2015}, who estimated 300 uninfected farms based on the crude density of farms in Durham County.

\begin{figure}
    \centering
    \includegraphics[width = .7\textwidth]{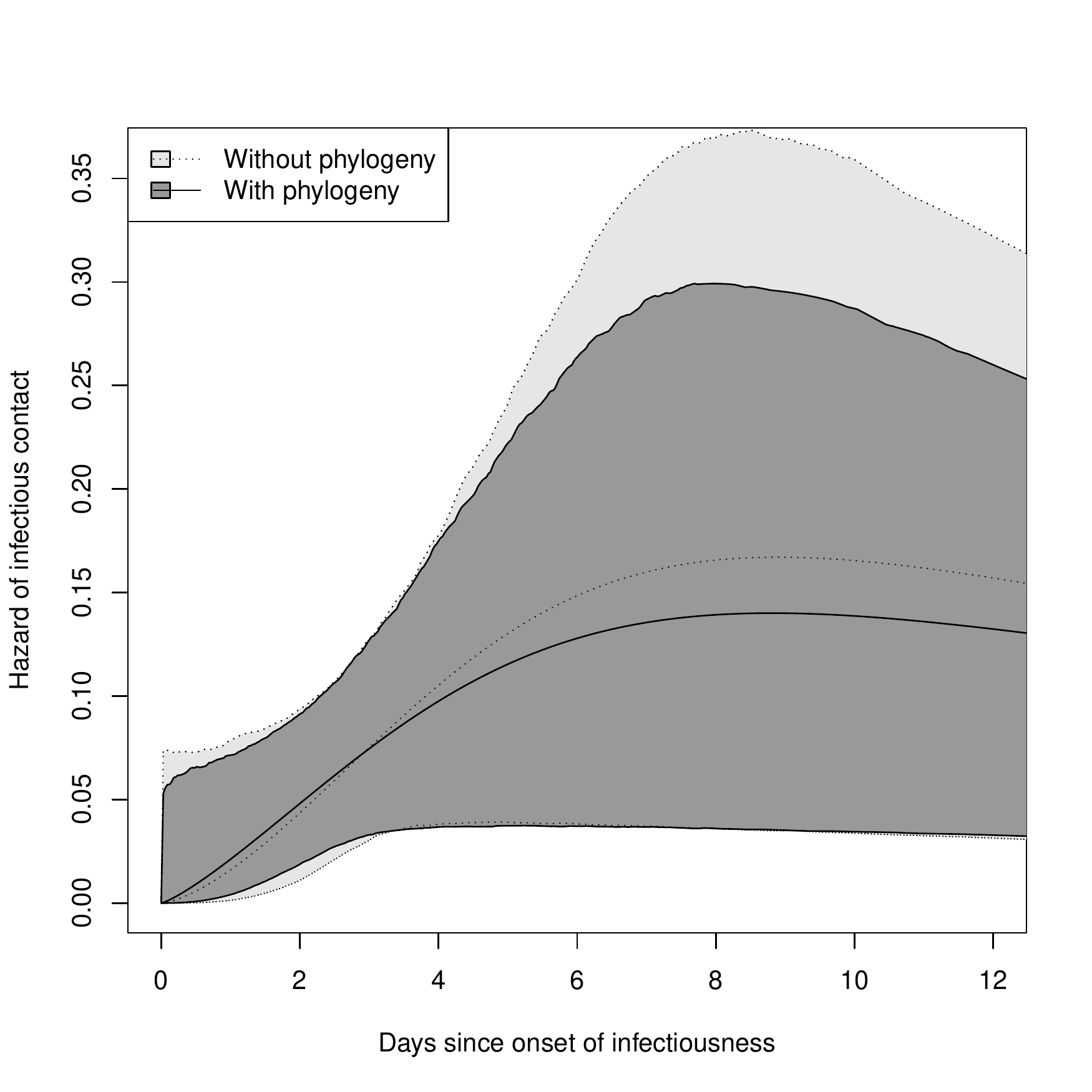}
    \caption{{\bf Point estimates and approximate 95\% confidence bands for the hazard function estimates.} These assume no farms escaped infection. The bands for the estimates using a phylogeny are narrower.}
    \label{fig:precision}
\end{figure}

To understand the effect of the phylogenies on the precision of the hazard function estimates, we constructed approximate pointwise 95\% confidence bands for the hazard functions estimated with no uninfected farms. We took 4000 samples from each likelihood using a grid with spacing $0.02$ on the interval $[-3.6, -1.3]$ for $\ln \lambda$ and $[-0.5, 1.8]$ for $\ln \gamma$. These intervals include the 99\% confidence limits for both estimates plus a boundary of width $0.2$. For each sample, we calculated the hazard function at 500 time points between 0 and 15. At each time point, we took the .025 and .975 quantiles of the calculated hazards to get an approximate 95\% confidence interval. Fig~\ref{fig:precision} shows that the confidence bands for the estimates with phylogenies are narrower. Similar results were obtained when parameters were sampled from their approximate multivariate normal distributions, which can be done using~\nameref{S3txt}.

\begin{table}
    \centering
    \caption{{\bf Log-logistic rate and shape parameter estimates.}}
    \begin{tabular}{cccc}
    \hline
    \textbf{Uninfected} & & \textbf{Without phylogeny} & \textbf{With phylogeny} \\
    \textbf{farms} & & Point estimate (95\% CI) & Point estimate (95\% CI) \\ \hline
    \multirow{2}{*}{0} & rate ($\lambda$) & 0.132 (0.062, 0.180) & 0.125 (0.060, 0.175) \\
    & shape ($\gamma$) & 2.486 (1.131, 4.748) & 2.233 (1.158, 3.807) \\[5pt]

    \multirow{2}{*}{6} & rate ($\lambda$) & 0.068 (0.018, 0.110) & 0.062 (0.018, 0.103) \\
    & shape ($\gamma$) & 1.972 (0.946, 3.605) & 1.812 (0.966, 2.979) \\[5pt]

    \multirow{2}{*}{12} & rate ($\lambda$) & 0.049 (0.010, 0.089) & 0.043 (0.010, 0.080) \\
    & shape ($\gamma$) & 1.878 (0.907, 3.413) & 1.725 (0.924, 2.823) \\[5pt]

    \multirow{2}{*}{24} & rate ($\lambda$) & 0.034 (0.005, 0.070) & 0.029 (0.005, 0.061) \\
    & shape ($\gamma$) & 1.815 (0.881, 3.289) & 1.666 (0.894, 2.720) \\
    \hline
    \end{tabular}
    \label{tab:hazard}
\end{table}

\section*{Discussion}
Here we took a single phylogeny, derived the set of possible transmission trees, and estimated transmission parameters using likelihoods that are sums over the possible transmission trees. By restricting the set of possible transmission trees, incorporating a phylogeny into the analysis produced more efficient estimates of infectiousness hazard ratios and the baseline hazard of infectious contact. The efficiency gain was largest for infectiousness hazard ratio estimates. The combination of survival analysis and algorithms linking phylogenies to transmission trees can be incorporated into analyses based on many different statistical and phylogenetic methods. More precise estimates of transmission parameters, not the transmission trees themselves, will inform public health responses to emerging infections. 

We assumed complete observation of infection times, latent periods, and infectious periods, so our methods need to be extended to account for missing data and phylogenetic uncertainty. Bayesian MCMC with data augmentation is a well-established method of handling missing data in infectious disease epidemiology~\cite{ONeillRoberts1999}. Combining this with a Bayesian MCMC for phylogeny reconstruction would allow our likelihoods to be integrated over both missing data and phylogenetic uncertainty~\cite{YangCME}. Standard phylogenetic reconstruction assumes a well-mixed population, which may not be a good approximation for pathogen populations partitioned among hosts, but Bayesian methods can reconcile the reconstruction of phylogenies with possible transmission trees~\cite{HallRambaut2015}. Extending these methods to nonparametric or semiparametric models of disease transmission will require iterative approaches to the assignment of probabilities to possible transmission trees~\cite{Kenah5, Kenah6}. 

Our methods can be extended to diseases with complex within-host and between-host dynamics by adapting the likelihood in equation~\eqref{eq:Lphy} or the algorithms linking phylogenies to transmission trees. For example, Assumption~\ref{assum:single} requires a strict transmission bottleneck. If multiple pathogen lineages can be transmitted when a new host is infected at time $t$, sequences sampled from the infectee could have an MRCA with a branching time before $t$. The nodes hosted by the infectee would no longer form a subtree of the phylogeny. This is similar to deep coalescence (lineage sorting), which causes gene trees and species trees to have different topologies. Deep coalescence would be most likely to occur in a disease with a wide transmission bottleneck, substantial within-host diversity, and little within-host evolution~\cite{Maddison1997}. Algorithm~\ref{alg:host} could be adapted by allowing the infector of an $i \in \mathcal{V}_j$ to be the host of an interior node of the phylogeny whose child is hosted by $j$. The likelihood in equation~\eqref{eq:Lphy} would have to include the likelihood contributions of these additional host assignments, including the within-host pathogen dynamics allowing two parallel strains to pass through $i$ to $j$. 

\paragraph{Simulations} 
The simulations suggest that a phylogeny can recover much of the information that would be obtained by observing who-infected-whom. Incorporating a phylogeny generated more precise estimates of $\beta_\text{inf}$ and $\ln \lambda_0$. This increase in efficiency remained when infectiousness varied over the course of the infectious period, as in the Weibull models. The simulations used only phylogenetic topologies and assumed that all within-host topologies were equally likely, limiting the ability of the phylogeny to constrain the set of possible transmission trees. Using branching times and more realistic models of within-host pathogen evolution would allow greater information about who-infected-whom to be extracted from a phylogeny. 

The simulations that excluded data on household members who escape infection showed that this information is critical to estimating $\beta_\text{inf}$, $\beta_\text{sus}$, and $\ln \lambda_0$ accurately. These individuals do not appear anywhere on the pathogen phylogeny, so this point has escaped the attention of many researchers working on incorporating phylogenetics into the analysis of infectious disease transmission data. Any analysis that excludes this data should have an explicit justification based on a complete-data model---for example, the initial spread of a mass-action epidemic can be analyzed without data on escapees~\cite{Kenah4}. In general, epidemiologic studies of emerging infections should be designed to capture information on individuals who were exposed to infection but not infected, which might justify greater emphasis on detailed studies of households, schools, or other settings with rapid transmission and a clearly defined population at risk.

\paragraph{Data analysis}
The data analysis showed that the increased precision found in the simulations can be obtained in practice. The incorporation of phylogenies allowed more precise estimates of the hazard of FMDV transmission from infected to susceptible farms. For simplicity, our analysis assumed that the times of infectiousness onset were accurately estimated. A data-augmented MCMC~\cite{ONeillRoberts1999} could be used to account for uncertainty in the onset of infectiousness, showing the importance of extending our methods to account for missing data. 

A more important limitation of this analysis was the lack of data on uninfected farms. The hazard function estimates were highly sensitive to the number of uninfected farms in the area where the cluster occurred. These data often go uncollected in outbreaks because their importance is unrecognized. This insight has important implications for the theory and practice of molecular infectious disease epidemiology.

\section*{Acknowledgements}
The authors thank participants in the RAPIDD-EPI Workshop on Survival Analysis and Phylogenetics in Infectious Disease Epidemiology at the University of Florida in January, 2013 for useful comments on this research. TB is grateful to the University of Florida for hosting him on a sabbatical. EK was supported by National Institute of Allergy and Infectious Diseases (NIAID) grant R00 AI095302. EK, MEH, and IML were supported by National Institute of General Medical Sciences (NIGMS) grant U54 GM111274 and NIAID grant R01 AI116770. The content is solely the responsibility of the authors and does not represent the official views of NIAID, NIGMS, or the National Institutes of Health.

\section*{Supporting Information}
\subsection*{S1 Appendix}
\label{S1app}
{\bf Proofs of the lemmas and theorems.}

\subsection*{S1 Data}
\label{S1dat}
{\bf Parameters, point estimates, and 95\% confidence limits from the simulations.} Used in~\nameref{S2txt}.

\subsection*{S2 Data}
\label{S2dat}
{\bf Parameters, point estimates, and 95\% confidence limits from the simulations using data on infecteds only.} Used in~\nameref{S2txt}.

\subsection*{S3 Data}
\label{S3dat}
{\bf Parameters, point estimates, and 95\% confidence limits from the simulations with Weibull contact intervals.} Used in~\nameref{S2txt}.

\subsection*{S1 Text}
\label{S1txt}
{\bf Python code for the simulations.}

\subsection*{S2 Text}
\label{S2txt}
{\bf R code for the simulation data analysis.} Including Tables~\ref{tab:mse}--\ref{tab:mse-Weibull} and Figs~\ref{fig:infRE}--\ref{fig:lnlambdaRE}.

\subsection*{S3 Text}
\label{S3txt}
{\bf R code for the FMDV data analysis.} Including Tab~\ref{tab:hazard} and Figs~\ref{fig:2001map}--\ref{fig:2001timeline} and \ref{fig:2001first}--\ref{fig:precision}.

\subsection*{S4 Text}
\label{S4txt}
{\bf Newick file for rescaled Durham cluster phylogeny}. Used in~\nameref{S3txt}.


\bibliographystyle{plos2015}
\bibliography{HSalgorithm}

\begin{thebibliography}{10}

\bibitem{Grenfell2004}
Grenfell BT, Pybus OG, Gog JR, Wood JLN, Daly JM, Mumford JA, et~al.
\newblock Unifying the epidemiological and evolutionary dynamics of pathogens.
\newblock Science. 2004;327:327--332.

\bibitem{Wilson2005}
Wilson DJ, Falush D, McVean G.
\newblock Germs, genomes and genealogies.
\newblock Trends in Ecology and Evolution. 2005;20:39--45.

\bibitem{Lemey2009}
Lemey P, Rambaut A, Drummond AJ, Suchard MA.
\newblock Bayesian phylogeography finds its roots.
\newblock PLoS Computational Biology. 2009;5:e1000520.

\bibitem{Rambaut2001}
Rambaut A, Robertson DL, Pybus OG, Peeters M, Holmes EC.
\newblock Phylogeny and the origin of {HIV}-1.
\newblock Nature. 2001;410:1047--1048.

\bibitem{Gilbert2007}
Gilbert MTP, Rambaut A, Wlasiuk G, Spira TJ, Pitchenik AE, Worobey M.
\newblock The emergence of {HIV/AIDS} in the {A}mericas and beyond.
\newblock PNAS. 2007;104:18566--18570.

\bibitem{Nelson2007}
Nelson MI, Simonesen L, Viboud C, Miller MA, Holmes EC.
\newblock Phylogenetic analysis reveals the global migration of seasonal
  influenza {A} viruses.
\newblock PLoS Pathogens. 2007;3:e131.

\bibitem{Rambaut2008}
Rambaut A, Pybus OG, Nelson MI, Viboud C, Taubenberger JK, Holmes EC.
\newblock The genomic and epidemiological dynamics of human influenza {A}
  virus.
\newblock Nature. 2008;453:615--619.

\bibitem{Russell2008}
Russell CA, Jones TC, Barr IG, Cox NJ, Garten RA, Gregory V, et~al.
\newblock The global circulation of seasonal influenza {A (H3N2)} viruses.
\newblock Science. 2008;320:340--346.

\bibitem{Bedford2015}
Bedford T, Riley S, Barr IG, Broor S, Chadha M, Cox NJ, et~al.
\newblock Global circulation patterns of seasonal influenza viruses vary with
  antigenic drift.
\newblock Nature. 2015;{DOI}:10.138/nature14460.

\bibitem{PybusRambaut2001}
Pybus OG, Rambaut A, Harvey PH.
\newblock An integrated framework for the inference of viral population history
  from reconstructed genealogies.
\newblock Genetics. 2000;155:1429--1437.

\bibitem{Pybus2001}
Pybus O, Charleston MA, Gupta S, Rambaut A, Holmes EC, Harvey PH.
\newblock The epidemic behavior of the hepatitis {C} virus.
\newblock Science. 2001;292:2323--2325.

\bibitem{Biek2007}
Biek R, Henderson JC, Waller LA, Rupprecht CE, Real LA.
\newblock A high-resolution genetic signature of demographic and spatial
  expansion in epizootic rabies virus.
\newblock PNAS. 2007;104:7993--7998.

\bibitem{Volz2009}
Volz EM, Kosakovsky~Pond SL, Ward MJ, Leigh~Brown AJ, Frost SDW.
\newblock Phylodynamics of infectious disease epidemics.
\newblock Genetics. 2009;183:1421--1430.

\bibitem{FrostVolz2010}
Frost SDW, Volz EM.
\newblock Viral phylodynamics and the search for an `effective number of
  infections'.
\newblock Philosophical Transactions of the Royal Society B.
  2010;365:1879--1890.

\bibitem{KoelleRasmussen2011}
Koelle K, Rasmussen DA.
\newblock Rates of coalescence for common epidemiological models at
  equilibrium.
\newblock Journal of the Royal Society Interface. 2011;9:997--1007.

\bibitem{Stadler2013}
Stadler T, K\"{u}hnert D, Bonhoeffer S, Drummond AJ.
\newblock Birth-death skyline plot reveals temporal changes of epidemic spread
  in {HIV} and hepatitis {C} virus ({HCV}).
\newblock Proceedings of the National Academy of Science. 2013;110:228--233.

\bibitem{Rasmussen2011}
Rasmussen DA, Ratmann O, Koelle K.
\newblock Inference for nonlinear epidemiological models using genealogies and
  time series.
\newblock {PLoS} Computational Biology. 2011;7:e1002136.

\bibitem{Rasmussen2014}
Rasmussen DA, Volz EM, Koelle K.
\newblock Phylodynamic inference for structured epidemiological models.
\newblock {PLoS} Computational Biology. 2014;10:e1003570.

\bibitem{Ou1992}
Ou CY, Ciesielski CA, Myers G, Bandea CI, Luo CC, Korber BTM, et~al.
\newblock Molecular epidemiology of {HIV} transmission in a dental practice.
\newblock Science. 1991;256:1165--1171.

\bibitem{Holmes1993}
Holmes EC, Zhang LQ, Simmonds P, Rogers AS, Leigh~Brown AJ.
\newblock Molecular investigation of {H}uman {I}mmunodeficiency {V}irus ({HIV})
  infection in a patient of an {HIV}-infected surgeon.
\newblock Journal of Infectious Diseases. 1993;167:1411--1414.

\bibitem{Leitner1996}
Leitner T, Escanilla D, Franz\'{e}n C, Uhl\'{e}n M, Albert J.
\newblock Accurate reconstruction of a known {HIV}-1 transmission history by
  phylogenetic tree analysis.
\newblock PNAS. 1996;93:10864--10869.

\bibitem{Leitner1999}
Leitner T, Albert J.
\newblock The molecular clock of {HIV} unveiled through analysis of a known
  transmission history.
\newblock PNAS. 1999;96:10752--10757.

\bibitem{Spada2004}
Spada E, Sagliocca L, Sourdis J, Garbuglia AR, Poggi V, De~Fusco C, et~al.
\newblock Use of the minimum spanning tree model for molecular epidemiological
  investigation of a nosocomial outbreak of hepatitis {C} virus infection.
\newblock Journal of Clinical Microbiology. 2004;42:4230--4236.

\bibitem{Jombart2011}
Jombart T, Eggo RM, Dodd PJ, Balloux F.
\newblock Reconstructing disease outbreaks from genetic data: a graph approach.
\newblock Heredity. 2011;106:383--390.

\bibitem{Snitkin2012}
Snitkin ES, Zelazny AM, Thomas PJ, Stock F, Program NCS, Henderson DK, et~al.
\newblock Tracking a hospital outbreak of carbapenem-resistant
  \emph{{K}lebsiella pneumoniae} with whole-genome sequencing.
\newblock Science Translational Medicine. 2012;4:148ra116.

\bibitem{Wertheim2011}
Wertheim JO, Kosakovsky~Pond SL, Little SJ, De~Gruttola V.
\newblock Using HIV transmission networks to investigate community effects in
  {HIV} prevention trials.
\newblock PLoS ONE. 2011;6:e27775.

\bibitem{Ypma2012}
Ypma RJF, Bataille AMA, Stegeman A, Koch G, Wallinga J, van Ballegooijen WM.
\newblock Unravelling transmission trees of infectious diseases by combining
  genetic and epidemiological data.
\newblock Proceedings of the Royal Society B. 2012;279:444--450.

\bibitem{Jombart2014}
Jombart T, Cori A, Didelot X, Cauchemez S, Fraser C, Ferguson N.
\newblock Bayesian reconstruction of disease outbreaks by combining
  epidemiologic and genomic data.
\newblock PLoS Computational Biology. 2014;10:e1003457.

\bibitem{Morelli2012}
Morelli MJ, Th\'{e}baud G, Chad\oe{}uf J, King DP, Haydon DT, Soubeyrand S.
\newblock A {B}ayesian inference framework to reconstruct transmission trees
  using epidemiological and genetic data.
\newblock PLoS Computational Biology. 2012;8:e1002768.

\bibitem{Cottam2006}
Cottam EM, Haydon DT, Paton DJ, Gloster J, Wilesmith JW, Ferris NP, et~al.
\newblock Molecular epidemiology of the foot-and-mouth disease virus outbreak
  in the {U}nited {K}ingdom in 2001.
\newblock Journal of Virology. 2006;80:11274--11282.

\bibitem{Cottam2008a}
Cottam EM, Th\'{e}baud G, Wadsworth J, Gloster J, Mansley L, Paton DJ, et~al.
\newblock Integrating genetic and epidemiological data to determine
  transmission pathways of foot-and-mouth disease virus.
\newblock Proceedings of the Royal Society B. 2008;275:887--895.

\bibitem{Cottam2008b}
Cottam EM, Wadsworth J, Shaw AE, Rowlands RJ, Goatley L, Maan S, et~al.
\newblock Transmission pathways of foot-and-mouth disease virus in the {U}nited
  {K}ingdom in 2007.
\newblock PLoS Pathogens. 2008;4:e1000050.

\bibitem{Gardy2011}
Gardy JL, Johnston JC, Ho~Sui SJ, Cook VJ, Shah L, Brodkin E, et~al.
\newblock Whole-genome sequencing and social-network analysis of a tuberculosis
  outbreak.
\newblock New England Journal of Medicine. 2011;364:730--739.

\bibitem{Didelot2012}
Didelot X, Eyre DW, Cule M, Ip CLC, Ansari MA, Griffiths D, et~al.
\newblock Microevolutionary analysis of \emph{Clostridium difficile} genomes to
  investigate transmission.
\newblock Genome Biology. 2012;13:R118.

\bibitem{Didelot2013}
Didelot X, Nell S, Yang I, Woltemate S, van~der Merwe S, Suerbaum S.
\newblock Genomic evolution and transmission of \emph{Helicobacter pylori} in
  two {S}outh {A}frican families.
\newblock PNAS. 2013;110:13880--13885.

\bibitem{Bryant2013}
Bryant JM, Sch\"{u}rch AC, van Deutekom H, Harris SF, de~Beer JL, de~Jager V,
  et~al.
\newblock Inferring patient to patient transmission of \textit{Mycobacterium
  tuberculosis} from whole genome sequencing data.
\newblock BMC Infectious Diseases. 2013;13:110.

\bibitem{Worby2014a}
Worby CJ, Lipsitch M, Hanage WP.
\newblock Within-host bacterial diversity hinders accurate reconstruction of
  transmission networks from genomic distance data.
\newblock PLoS Computational Biology. 2014;10:e1003549.

\bibitem{Worby2014b}
Worby CJ, Chang HH, Hanage WP, Lipsitch M.
\newblock The distribution of pairwise genetic distances: A tool for
  investigating disease transmission.
\newblock Genetics. 2014;198:1395--1404.

\bibitem{PybusRambaut2009}
Pybus OG, Rambaut A.
\newblock Evolutionary analysis of the dynamics of viral infectious disease.
\newblock Nature Reviews Genetics. 2009;10:540--550.

\bibitem{Ypma2013-Genetics}
Ypma RJF, van Ballegooijen WM, Wallinga J.
\newblock Relating phylogenetic trees to transmission trees of infectious
  disease outbreaks.
\newblock Genetics. 2013;195:1055--1062.

\bibitem{RomeroSeverson2014}
Romero-Severson E, Skar H, Bulla I, Albert J, Leitner T.
\newblock Timing and order of transmission events is not directly reflected in
  a pathogen phylogeny.
\newblock Molecular Biology and Evolution. 2014;31:2472--2482.

\bibitem{Didelot2014}
Didelot X, Gardy J, Colijn C.
\newblock Bayesian inference of infectious disease transmission from
  whole-genome sequence data.
\newblock Molecular Biology and Evolution. 2014;31:1869--1879.

\bibitem{LauGibson2015}
Lau MSY, Marion G, Streftaris G, Gibson G.
\newblock A systematic {B}ayesian integration of epidemiological and genetic
  data.
\newblock PLoS Computational Biology. 2015;11:e1004633.

\bibitem{Kenah4}
Kenah E.
\newblock Contact intervals, survival analysis of epidemic data, and estimation
  of ${R}_0$.
\newblock Biostatistics. 2011;12:548--566.

\bibitem{Kenah5}
Kenah E.
\newblock Nonparametric survival analysis of epidemic data.
\newblock Journal of the Royal Statistical Society, Series B. 2013;75:277--303.

\bibitem{Kenah6}
Kenah E.
\newblock Semiparametric relative-risk regression for infectious disease
  transmission data.
\newblock Journal of the American Statistical Association. 2015;110:313--325.

\bibitem{RampeyLongini}
Rampey AH Jr, Longini IM Jr, Haber M, Monto AS.
\newblock A discrete-time model for the statistical analysis of infectious
  disease incidence data.
\newblock Biometrics. 1992;48:117--128.

\bibitem{Kenah3}
Kenah E, Lipsitch M, Robins JM.
\newblock Generation interval contraction and epidemic data analysis.
\newblock Mathematical Biosciences. 2008;213:71--79.

\bibitem{HallRambaut2015}
Hall M, Rambaut A.
\newblock Epidemic reconstruction in a phylogenetics framework: transmission
  trees as partitions of the node set.
\newblock PLoS Computational Biology. 2015;11:e1004613.

\bibitem{Sankoff1975}
Sankoff D.
\newblock Minimal mutation trees of sequences.
\newblock SIAM Journal of Applied Mathematics. 1975;28:35--42.

\bibitem{FMDV2014}
Spickler AR.
\newblock Foot and Mouth Disease.
\newblock Iowa State University Center for Food Security \& Public Health;
  April, 2014.

\bibitem{Gouy2010}
Gouy M, Guindon S, Gascuel O.
\newblock Seaview version 4: A multiplatform graphical user interface for
  sequence alignment and phylogenetic tree building.
\newblock Molecular Biology and Evolution. 2010;27:221--224.

\bibitem{ONeillRoberts1999}
O'Neill PD, Roberts GO.
\newblock Bayesian inference for partially observed stochastic epidemics.
\newblock Journal of the Royal Statistical Society, Series A.
  1999;162:121--129.

\bibitem{YangCME}
Yang Z.
\newblock Computational Molecular Evolution.
\newblock Oxford Series in Ecology and Evolution. Oxford: Oxford University
  Press; 2006.

\bibitem{Maddison1997}
Maddison WP.
\newblock Gene trees in species trees.
\newblock Systematic Biology. 1997;46:523--536.

\end{thebibliography}

\end{document}